\documentclass[10pt, journal]{IEEEtran}

\usepackage{layout}
\usepackage{graphicx}
\usepackage{amscd}

\usepackage{multirow}
\usepackage{amsmath}
\usepackage{array}
\usepackage{xy}
\newcommand{\mr}[1]{\multirow{2}*{#1}}

\usepackage{setspace}

\newtheorem{thm}{Theorem}[section]
\newtheorem{lem}[thm]{Lemma}

\newtheorem{cor}[thm]{Corollary}

\begin{document}

\title{Modeling Spatial and Temporal Dependencies of User Mobility in Wireless Mobile Networks}

\author{\authorblockN{Wei-Jen Hsu\authorrefmark{1},
Thrasyvoulos Spyropoulos\authorrefmark{2},
Konstantinos Psounis\authorrefmark{3} and
Ahmed Helmy\authorrefmark{1}} \\
\authorblockA{\authorrefmark{1}Dept. of Computer and Information Science and Engineering,
University of Florida,
Gainesville, Florida \\
\authorrefmark{2}Computer Engineering and Networks Lab, ETH Zurich\\
\authorrefmark{3}Dept. of Electrical Engineering,
University of Southern California,
Los Angeles, California
\\Email: wjhsu@ufl.edu,   spyropoulos@tik.ee.ethz.ch,  kpsounis@usc.edu,  helmy@ufl.edu}
}


\maketitle
\bibliographystyle{IEEEtran}

\begin{abstract}
Realistic mobility models are fundamental to evaluate the performance of protocols in mobile ad hoc networks. Unfortunately, there are no mobility models that capture the non-homogeneous behaviors in both space and time commonly found in reality, while at the same time being easy to use and analyze. Motivated by this, we propose a time-variant community mobility model, referred to as the TVC model, which realistically captures spatial and temporal correlations. We devise the communities that lead to skewed location visiting preferences, and time periods that allow us to model time dependent behaviors and periodic re-appearances of nodes at specific locations.

To demonstrate the power and flexibility of the TVC model, we use it to generate synthetic traces that match the characteristics of a number of qualitatively different mobility traces, including wireless LAN traces, vehicular mobility traces, and human encounter traces. More importantly, we show that, despite the high level of realism achieved, our TVC model is still theoretically tractable. To establish this, we derive a number of important quantities related to protocol performance, such as the average node degree, the hitting time, and the meeting time, and provide examples of how to utilize this theory to guide design decisions in routing protocols.
\end{abstract}

\section{Introduction}


{\it Mobile ad hoc networks} (MANETs) are self-organized, infrastructure-less networks that could potentially support many applications, such as vehicular networking (VANET)~\cite{DieselNet}, wild-life tracking~\cite{ZebraNet}, and Internet provision to rural areas~\cite{DTN-Fall}, to name a few. Mobility also enables message delivery in sparsely connected networks, generally known as delay tolerant networks (DTNs). As the devices are easily portable and the scenarios of deployment are inherently dynamic, {\it mobility} becomes one of the key characteristics in most of these networks. It has been shown that mobility impacts MANETs in multiple ways, such as network capacity~\cite{capacity-tse}, routing performance~\cite{IMPORTANT}, and cluster maintenance~\cite{cluster}. In short, the evaluation of protocols and services for MANETs seems to be inseparable from the underlying mobility models. It is, thus, of crucial importance to have suitable mobility models as the foundation for the study of ad hoc networks.

Ideally, a good mobility model should achieve a number of goals: (i) it should first capture {\it realistic} mobility patterns of scenarios in which one wants to eventually operate the network; (ii) at the same time it is desirable that the model is {\it mathematically tractable}; this is very important to allow researchers to derive performance bounds and understand the limitations of various protocols under the given scenario, as in \cite{single-copy, multi-copy, capacity-tse, Cambridge-trace}; (iii) finally, it should be {\it flexible} enough to provide qualitatively and quantitatively different mobility characteristics by changing some parameters of the model, yet in a repeatable and scalable manner; designing a new mobility model for each existing or new scenario is undesirable.

Most existing mobility models excel in one or, less often, two aspects of the above requirements, but none satisfies all of them at the same time. Our goal in this paper is, on one hand, to improve the existing {\it random mobility models} (e.g., random walk, random direction, etc.) and {\it synthetic mobility models} (e.g., \cite{WWP, RPGM, pathway-obstacle}) on the front of {\it realism}, by considering empirically observed mobility characteristics from the traces \cite{individual-study}. On the other hand, the construction of the model should new model should be simple enough to allow in-depth {\it theoretical analysis}, and be {\it flexible} enough to have wider applicability than the {\it mobility traces} (which provide only a single snapshot of the underlying mobility process) and current {\it trace-based mobility models} \cite{WLAN-model, T++-model, Dart-INFOCOM} which focus mainly on matching mobility characteristics with a specific class of traces.


The main contribution of this paper is the proposal of a {\it time-variant community mobility model}, referred to as the TVC model, which is {\it realistic, flexible, and mathematically tractable}. One salient characteristic in the TVC model is \emph{location preference}.  Another important characteristic is the \emph{time-dependent, periodical behavior} of nodes. To our best knowledge, this is the first {\it synthetic mobility model} that captures non-homogeneous behavior in both {\it space} and {\it time}.


To establish the flexibility of our TVC model we show that we can match its two prominent properties, {\it location visiting preferences} and {\it periodical re-appearance}, with {\it multiple} WLAN traces, collected from environments such as university campuses~\cite{Dart-trace, individual-study} and corporate buildings~\cite{MIT-trace}. More interestingly, although we motivate the TVC model with the observations made on WLAN traces, our model is generic enough to have wider applicability. We validate this claim by examples of matching our TVC model with two additional mobility traces: a vehicle mobility trace\cite{cab-spotting} and a human encounter trace\cite{Cambridge-trace}. In the latter case, we are even able to match our TVC model with some other mobility characteristics not explicitly incorporated in our model by its construction, namely the inter meeting time and encounter duration between different users/devices.

Finally, in addition to the improved realism, the TVC model can be mathematically treated to derive analytical expressions for important quantities of interest, such as the {\it average node degree}, the {\it hitting time} and the {\it meeting time}. These quantities are often fundamental to theoretically study issues such as routing performance, capacity, connectivity, etc. We show that our theoretical derivations are accurate through simulation cases with a wide range of parameter sets, and additionally provide examples of how our theory could be utilized in actual protocol design. To our best knowledge, this is the first synthetic mobility model proposed that matches with traces from multiple scenarios, and has also been theoretically treated to the extent presented in this paper. We make the code of the TVC model available at~\cite{simu_code_download}.

The the paper is organized as follows: In Section \ref{RW} we discuss related work. Our {\it TVC model} is then introduced in Section \ref{model-description}. In Section \ref{match_trace}, we show how to generate realistic mobility scenarios matched with various traces. Then, in Section \ref{theory}, we present our theoretical framework and derive generic expressions of various quantities. Simulation validates the accuracy of these expressions in Section \ref{match_theory_sim}. Additionally, in Section \ref{perf_prediction}, we motivate our theoretical framework further, by applying our analysis to performance predictions in protocol design. Finally, we conclude the paper in Section \ref{conclusion}.

\section{Related Work} \label{RW}

Mobility models have been long recognized as one of the fundamental components that impacts the performance of wireless ad hoc networks. A wide variety of mobility models are available in the research community (see \cite{mobility-model} for a good survey). Among all mobility models, the popularity of {\it random mobility models} (e.g., random walk, random direction, and random waypoint) roots in its simplicity and mathematical tractability. A number of important properties for these models have been studied, such as the stationary nodal distribution~\cite{random-waypoint}, the hitting and meeting times~\cite{Akis-MOBIHOC}, and the meeting duration~\cite{apoorva-theory}. These quantities in turn enable routing protocol analysis to produce performance bounds~\cite{single-copy, multi-copy}. However, {\it random mobility models} are based on over-simplified assumptions, and as has been shown recently and we will also show in the paper, the resulting mobility characteristics are very different from real-life scenarios. Hence, it is debatable whether the findings under these models will directly translate into performance in real-world implementations of MANETs.

More recently, an array of {\it synthetic mobility models} are proposed to improve the realism of the simple {\it random mobility models}. More complex rules are introduced to make the nodes follow a popularity distribution when selecting the next destination~\cite{WWP}, stay on designated paths for movements~\cite{pathway-obstacle}, or move as a group~\cite{RPGM}. These rules enrich the scenarios covered by the {\it synthetic mobility models}, but at the same time make theoretical treatment of these models difficult. In addition, most {\it synthetic mobility models} are still limited to {\it i.i.d.} models, and the mobility decisions are also independent of the current location of nodes and time of simulation.

A different approach to mobility modeling is by {\it empirical mobility trace collection}. Along this line, researchers have exploited existing wireless network infrastructure, such as wireless LANs (e.g., \cite{MIT-trace, UCSD-trace, Dart-trace}) or cellular phone networks (e.g., \cite{reality-mining}), to track user mobility by monitoring their locations. Such traces can be replayed as input mobility patterns for simulations of network protocols~\cite{group-study}. More recently, DTN-specific testbeds~\cite{Cambridge-trace, DieselNet, ZebraNet} aim at collecting encounter events between mobile nodes instead of the mobility patterns. Some initial efforts to mathematically analyze these traces can be found in~\cite{Cambridge-trace,Karagiannis:Mobicom07}. Yet, the size of the traces and the environments in which the experiments are performed can not be adjusted at will by the researchers. To improve the flexibility of traces, the approach of {\it trace-based mobility models} have also been proposed~\cite{WLAN-model, T++-model, Dart-INFOCOM}. These models discover the underlying mobility rules that lead to the observed properties (such as the duration of stay at locations, the arrival patterns, etc.) in the traces. Statistical analysis is then used to determine proper parameters of the model to match it with the particular trace.

The goal of this work is to combine the strengths of various approaches to mobility modeling and propose a {\it realistic, flexible, and mathematically tractable synthetic mobility model}. Our work is partly motivated by several prominent, common properties in multiple WLAN traces (e.g., traces available from public archives~\cite{MobiLib-web, CRAWDAD-web}) we observed in \cite{individual-study}, based on which we construct the TVC model. This model extends the concept of communities proposed by us in~\cite{Akis-MOBIHOC} and also introduces time-dependent behavior. A preliminary version of the model has been presented in~\cite{conf-version-TVC}. In this work we highlight the {\it flexibility} of the TVC model by matching the synthetic traces with two additional, qualitatively different traces to WLAN traces (i.e., vehicular and human encounter traces, in section \ref{match_trace}). We also extend and present more generic theoretical results under the scenario with multiple communities (section \ref{theory}), and display its applications on protocol performance prediction (section \ref{perf_prediction}).

We differentiate our work from other trace-based models~\cite{WLAN-model, T++-model, Dart-INFOCOM} in several aspects. First, among all efforts of providing realistic mobility models, to our best knowledge, this is the first work to explicitly capture time-variant mobility characteristics. Although capturing time-dependent behavior is suggested in~\cite{Dart-INFOCOM}, it has not been incorporated in the particular paper. Second, while previous works emphasize the capability to truthfully recreate the mobility characteristics observed from the traces, we also strive to ensure at the same time the mathematical tractability of the model. Our motivation is to facilitate the application of our model for performance prediction of various communication protocols. Finally,
most of the other trace-based models have not been shown as capable to match mobility characteristics of a diverse set of traces, since their focus is mostly on one particular trace or at most a single class of traces (e.g., WLAN trace). We go beyond that and re-produce matching mobility characteristics of several {\it qualitatively different} traces, including WLAN, vehicle, and human encounter traces.


As a final note, in \cite{community-network}, the authors assume the attraction of a community (i.e., a geographical area) to a mobile node is derived from the number of friends of this node currently residing in the community. In our paper we assume that the nodes make movement decisions independently of the others (nonetheless, node sharing the same community will exhibit mobility correlation, capturing the social feature indirectly). Mobility models with inter-node dependency require a solid understanding of the social network structure, which is an important area under development. We plan to work further in this direction in the future.

\section{Time-variant Community Mobility Model} \label{model-description}

\subsection{Mobility Characteristics Observed in WLAN Traces} \label{observation}

The main objective of this paper is to propose a mobility model that captures the important mobility characteristics observed in daily life. To better understand this mobility, we have conducted extensive analysis of a number of wireless LAN traces collected by several research groups (e.g., traces available at \cite{MobiLib-web} or \cite{CRAWDAD-web}). The reason for this choice is that WLAN traces log information regarding large numbers of nodes, and thus are reliable for statistical analysis. After analyzing a large number of traces, we have observed two important properties that are common in all of them: (a){\it skewed location visiting preferences} and (b){\it time-dependent mobility behavior}~\cite{individual-study}.

\begin{figure}

\begin{minipage}[t]{1.7in}

\centering
\includegraphics[height=1.5in, angle=-90]{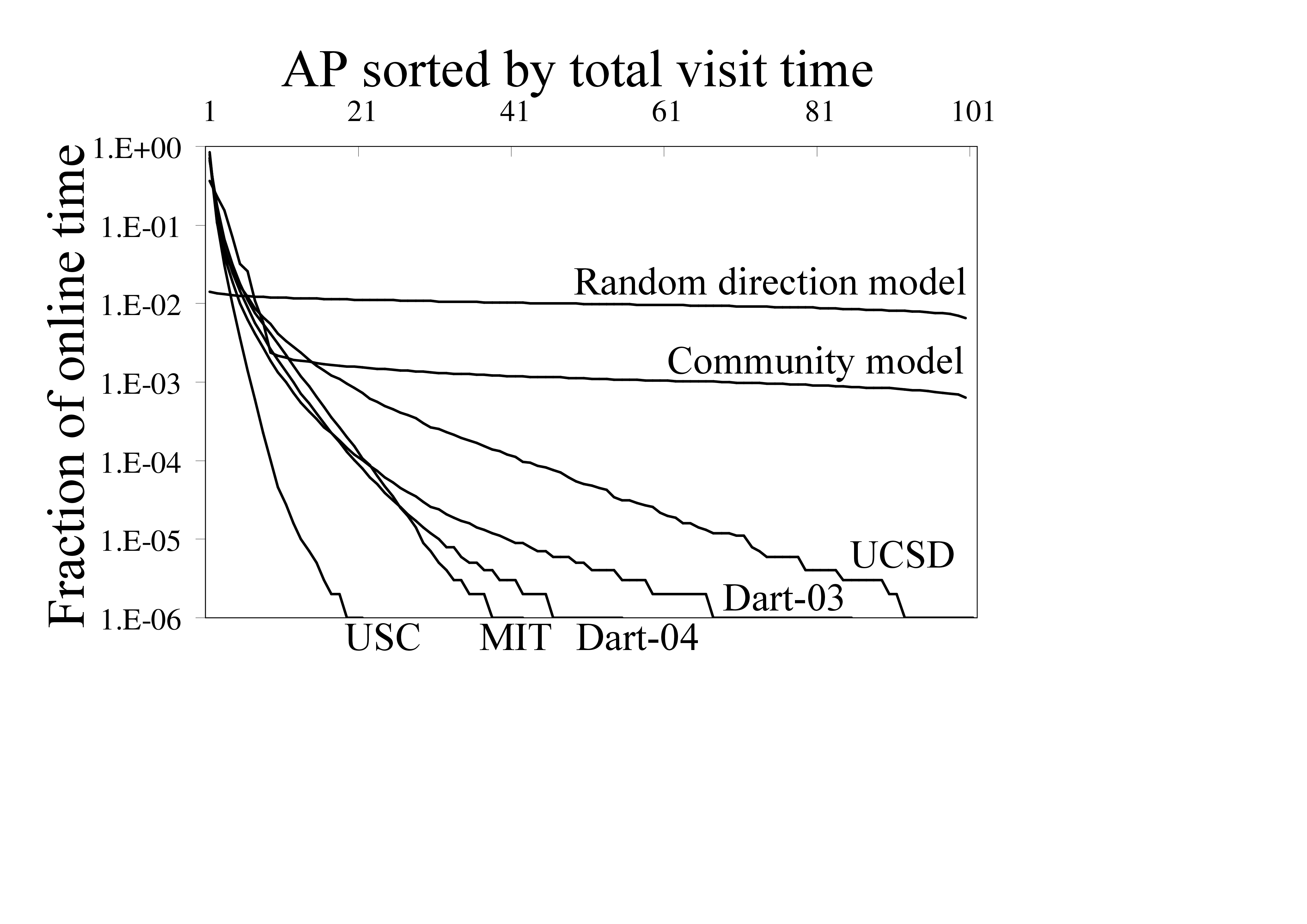}

\vspace{0.05in}

\footnotesize{(a) Skewed location visiting preferences.}

\end{minipage}
\hfill
\begin{minipage}[t]{1.7in}

\centering
\includegraphics[height=1.5in, angle=-90]{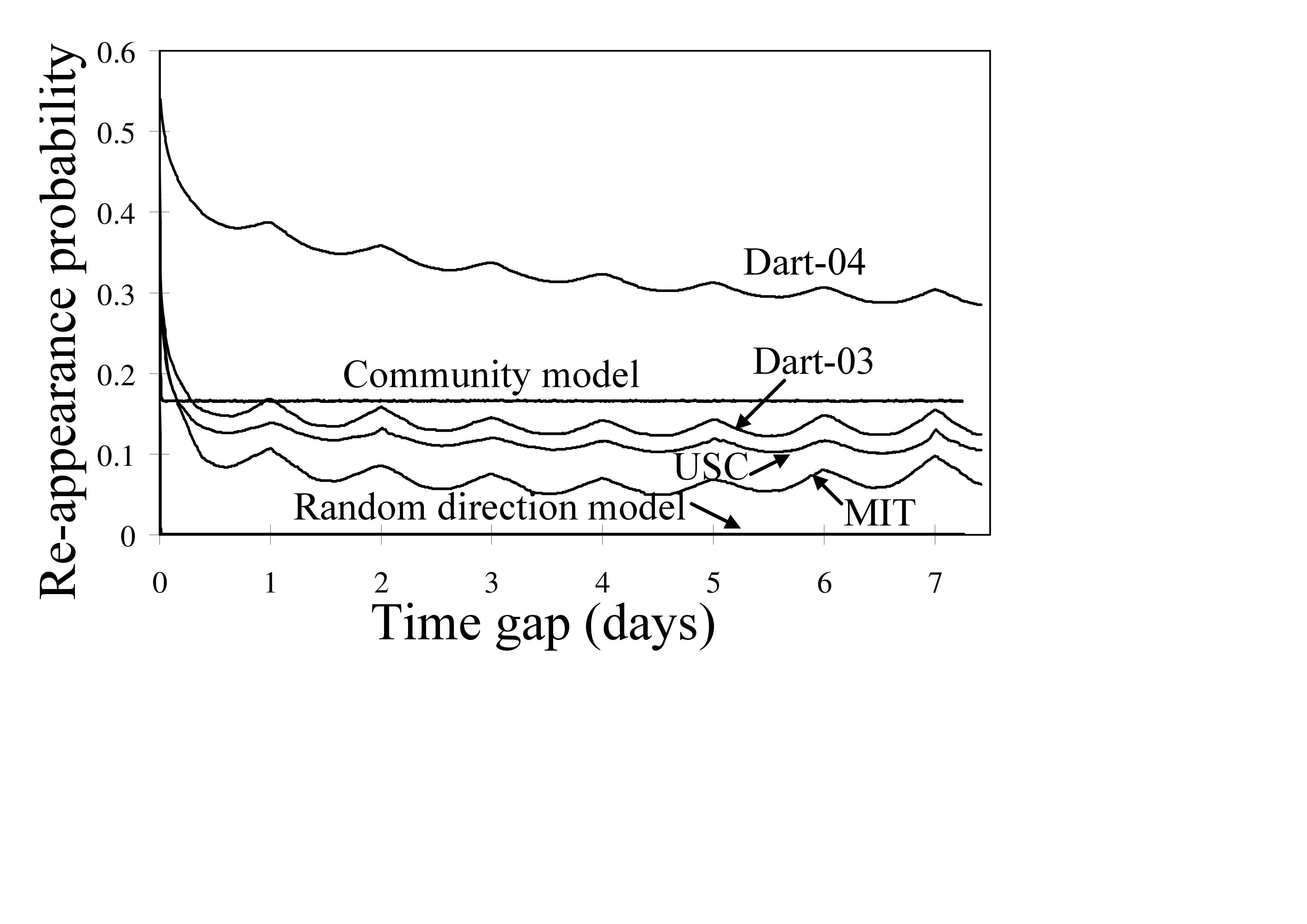}

\vspace{0.05in}

\footnotesize{(b) Periodical re-appearance at the same location.}

\end{minipage}
\hfill

\caption{Two important mobility features observed from WLAN traces. Labels of traces used: MIT: trace from \cite{MIT-trace}, Dart: trace from \cite{Dart-trace}, UCSD: trace from \cite{UCSD-trace}, USC: trace from \cite{individual-study}.}
\label{pref-and-reappear}

\end{figure}

More specifically, the \emph{location visiting preference} refers to the percentage of time a node spends at a given access point (AP). We refer to the coverage area of an access point as a {\it location}. In Fig. \ref{pref-and-reappear}(a), we draw the probability density function of the percentages of online time an average user spends at each location, ranking the locations from the most favorite place to the least for various traces. The distribution appears highly {\it skewed}; more than $95\%$ of user's online time is spent at only top five APs. The \emph{time-dependent mobility behavior} refers to the observation that nodes visit different locations, depending the time of the day. In Fig. \ref{pref-and-reappear}(b) we plot the probability of a node re-appearing at the same location at some time in the future, as a function of the elapsed time. It is clear that this probability displays some amount of periodicity, as the mobile nodes have stronger tendency to re-appear at a previously visited location after a time gap of integer multiples of days. A slightly higher peak on the 7th day, suggesting a stronger weekly correlation in location visiting preferences, could also be observed in some curves (e.g., MIT).

Unfortunately, these two prominent realistic mobility characteristics are not captured by commonly used simple random models, as they do not possess any space or time dependent features in user mobility. This is demonstrated in Fig.~\ref{pref-and-reappear} by a straight line (uniform distribution) for the Random Direction model. The same could be obtained from Random Waypoint, Random walk, etc., or even more sophisticated models without spatial-temporal preferences (e.g., \cite{RPGM, pathway-obstacle}). There are some more recent models (e.g., \cite{Akis-MOBIHOC, WWP, WLAN-model, T++-model}) that aim at capturing spatial preference explicitly. As shown in Fig. \ref{pref-and-reappear}(a) using the simple community model~\cite{Akis-MOBIHOC}, with appropriately assigned parameters this model is able to capture the {\it skewed location visiting preference}, to some extent. However, time-dependent behavior is not captured, and thus the {\it periodical re-appearance} property cannot be reproduced, as shown by the flat curve labeled {\it community model} in Fig. \ref{pref-and-reappear}(b).

It is our goal to design a mobility model that successfully captures the {\it skewed location preference} and {\it time-dependency} mobility properties observed in the traces {\it in an analytically tractable fashion}. We believe that although the above observations are made based on WLAN traces, the two properties in question are indeed prevalent in real-life mobility. This belief is supported by typical daily activities of humans: most of us tend to spend most time at a handful of frequently visited locations, and a recurrent daily or weekly schedule is an inseparable part of our lives. It is essential to design a model that captures such spatial-temporal preferences of human mobility in many contexts.

\subsection{Construction of the Time-variant Community Model} \label{model-spec}

\begin{table}
\caption{Parameters of the time-variant community mobility model$^{1}$ }
\scriptsize
\label{parameter_table}


\begin{center}
\begin{tabular}{|c||c|}
\hline
$N$ & Edge length of simulation area \\
\hline
$ V $ & Number of time periods \\
\hline
$T^t $  & Duration of $t$-th time period\\
\hline
$S^t $ & Number of communities in time period $t$\\
\hline
$C_j^t$ & Edge length of community $j$ in time period $t$\\
\hline
$Comm_j^t$ & The $j$-th community during time period $t$ \\
\hline
\mr{$p_{i,j}^t$} & The probability to choose community $j$ when\\
& the previous community is $i$, during time period $t$ \\
\hline
\mr{$\pi_{j}^t$} & Stationary probability of an epoch in  \\
& community $j$ during time period $t$\\
\hline
$v_{min}, v_{max}, \overline{v}$ & Minimum, maximum, and average speed\footnotemark[1] \\
\hline
$D_{max,j}, \overline{D_j}$ & Maximum and average pause time after each epoch\footnotemark[1] \\
\hline
$\overline{L_j}$ & Average epoch length for community $j$\\
\hline
\mr{$P_{move,j}^{t} | P_{pause,j}^{t}$} & Probability that a node is moving $|$ pausing\\
&  when being in community $j$ during period $t$ \\
\hline
\mr{$P_j^t$} & Fraction of time the node is in \\
& state $j$ ($P_j^t = P_{move,j}^t +P_{pause,j}^t$) \\
\hline
$K$ & Transmission range of nodes \\
\hline
\mr{$A(a_j^t, b_k^t)$} & The overlapped area between $Comm_j^t$ of node $a$ \\
& and $Comm_k^t$ of node $b$ \\
\hline
\mr{$w^t$} & A specific relationship between a target coordinate \\
 & and the communities in time period $t$\\
\hline
\mr{$\Omega^t$} & The set of all possible relationships between \\
 & a target coordinate and the communities in time period $t$\\
\hline
$P_{h} (w^t)$ & Unit-time hitting probability \\
 & under the specific scenario $w^t$\\
\hline
$P_{H} (w^t)$ & Hitting probability for a time period $t$\\
 & under specific scenario $w^t$\\
\hline
$P_{m}^t$ & Unit-time meeting probability in time period $t$\\
\hline
$P_{M}^t$ & Meeting probability for a time period $t$\\
\hline

\end{tabular}
\end{center}
\end{table}

\begin{figure}
\centering
\includegraphics[width=3.0in]{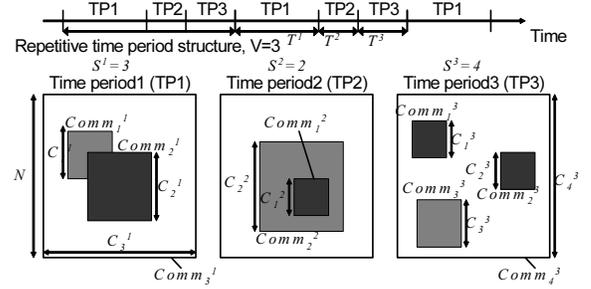}

\caption{Illustration of a generic scenario of time-variant mobility model, with three time periods and different numbers of communities in each time period.}
\label{comm-tp-illustration}
\end{figure}

In this section, we present the design of our {\it time-variant community (TVC) mobility model}. We illustrate the model with an example in Fig. \ref{comm-tp-illustration} and use this example to introduce the notations we use (see Table \ref{parameter_table}) in the rest of the paper. 


First, to induce {\it skewed location visiting preferences}, we define some {\it communities} (or heavily-visited geographic areas). Take time period 1 (TP1) in Fig. \ref{comm-tp-illustration} as an example, the communities are denoted as $Comm_j^1$ and each of them is a square geographical area with edge length $C_j^1$.\footnote{For all parameters used in the paper, we follow the convention that the subscript of a quantity represents its community index, and the superscript represents the time period index.} A node visits these communities with different {\it probabilities} (details are given later) to capture its spatial preference in mobility. In the TVC model, the mobility process of a node consists of {\it epochs} in these communities. When the node chooses to have an {\it epoch} in community $j$ (we say that  {\it the node is in state $j$} during this epoch), it starts from the end point of the previous epoch within $Comm_j^1$ and the epoch length (movement distance) is drawn from an exponential distribution with average $\overline{L_j}$, in the same order of the community edge length. The node then picks a random speed uniformly in $[v_{min}$, $v_{max}]$, and a direction (angle) uniformly in $[0, 2\pi]$, and performs a random direction movement within the chosen community with the chosen epoch length\footnote{To avoid boundary effects, if the node hits the community boundary it is re-inserted from the other end of the area (i.e., "torus" boundaries). Note that we could also choose random waypoint or random walk models for the type of movement during each epoch.}. The first difference between the TVC model and the standard Random Direction model is hence the spatial preference and location-dependent behavior. Note that, a node can still roam around the whole simulation area during some epochs, by assigning an additional community that corresponds to the whole simulation field (e.g. $Comm_{3}^{1}$). We refer to such epochs as {\it roaming epochs}.

We next explain how a node selects the next community for a sequence of epochs. At the completion of an epoch, the node remains stationary for a pause time uniformly chosen in $[0,D_{max,j}]$. Then, depending on its current state $i$ and time period $t$, the node chooses the next epoch to be in community $j$ with probability $p_{i,j}^t$. This community selection process is essentially a time-variant Markov chain that captures the spatial and temporal dependencies in nodal mobility and thus makes the community selection process in the TVC model non-{\it i.i.d.}, an important feature absent in many synthetic mobility models even if they consider non-uniform mobility features. Now, if the end point of the previous epoch is in $Comm_j^t$ (this can be the case when the node has two consecutive epochs in $Comm_j^t$, or $Comm_j^t$ contains $Comm_i^t$), the node starts the next epoch directly. If, on the other hand, the node is currently not in $Comm_j^t$, a {\it transitional epoch} is inserted to bridge the two epochs in disjoint communities. The node selects a random coordinate point in the next community, moves directly towards this point on the shortest straight path with a random speed drawn from $[v_{min}$, $v_{max}]$, and then continues with an epoch in the next community. Hence the movement trajectory of a node is always continuous in space.

\begin{figure}
\centering
\includegraphics[width=3.0in]{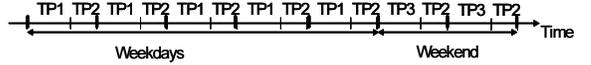}

\caption{An illustration of a simple weekly schedule, where we use time period $1$ (TP1) to capture weekday working hour, TP2 to capture night time, and TP3 to capture weekend day time.}
\label{weekly-schedule}
\end{figure}

We next introduce the structure in time. To capture time-dependent behavior, one creates multiple {\it time periods} with different community and parameter settings. As an example, there are $V=3$ time periods with duration $T^1$, $T^2$, and  $T^3$ in Fig. \ref{comm-tp-illustration}. These time periods follow a {\it periodic structure} (e.g., a simple recurrent structure in Fig.~\ref{comm-tp-illustration} or the weekly schedule in Fig. \ref{weekly-schedule}). This setup naturally captures the {\it temporal preferences} (e.g., go to work during the days and home during the nights) and {\it periodicity} in human mobility. On the time boundaries between time periods, each node continues with its ongoing epoch, and decides the next epoch according to the new parameter settings in the new time period when it finishes the current epoch.

As a final note, we choose to construct the TVC model with simple building blocks introduced above due to its amenability to theoretical analysis~\cite{Akis-MOBIHOC} and flexibility. To further explain the flexibility of our TVC model, we note that the number of communities in each time period (denoted as $S^t$) can be different, and the communities can overlap (as in TP1 in Fig. \ref{comm-tp-illustration}) or contain each other (as in TP2 in Fig. \ref{comm-tp-illustration}). Finally, the time period structure, communities, and all other parameters could be assigned differently for {\it each node} to capture node-dependent mobility (e.g., people following different schedules, with different working places, etc.), while nodes can share some communities (i.e., the popular locations) as well. This construction allows for maximum flexibility when setting up the simulations for nodes with heterogeneous behaviors\footnote{When necessary, we use a pair of parentheses to include the node ID for a particular parameter, e.g., $C_j^t (i)$ denotes the edge length of the $j$-th community during time period $t$ for node $i$.}.

The benefit of using simple building blocks will become evident in Section~\ref{theory}. At the same time, we will show next that these choices do not compromise our model's ability to accurately capture real life mobility scenarios.

\section{Generation of Mobility Scenarios} \label{match_trace}

The TVC model described in the previous section provides a general framework to model a wide range of mobility scenarios. In this section, our aim is to demonstrate the model's flexibility and validate its realism by generating various synthetic traces from the model, with matching mobility characteristics to well-known, publicly-available traces (e.g., WLANs, VANET, and human encounter traces). However, it is important to note that the use of such a model is not merely to match it with any specific trace instance available; this is only done for validation and calibration purposes. Rather, the goal is to be able to reproduce a much larger range of realistic mobility instances than a single trace can provide\footnote{We have made our mobility trace generator available at \cite{simu_code_download}. The tool provides mobility traces in both ns-2\cite{ns-2} compatible format and time-location (i.e., $(t, x, y)$) format.}.


We first outline a general 3-step systematic process to construct specific mobility scenarios. Then, we demonstrate our success to generate matching mobility characteristics with three qualitatively different traces. All the parameter values we use in this section are also available in \cite{simu_code_download}\footnote{Due to space limitations, we cannot list all parameters in this paper.}.

\noindent \textbf{\underline{STEP 1: Determine the Structure in Space and Time}}


\noindent $\bullet$ (1.1 Number of communities) Each community in the TVC model corresponds to a location visited frequently by nodes (i.e., the most visited location in Fig. \ref{pref-and-reappear}(a) corresponds to the most popular community in the model, and so on). The number of communities needed is thus determined by how closely one wants the mobility characteristics to match with the curves in Fig. \ref{pref-and-reappear}(a). Due to the nature of skewed location visiting preference, in our experience, only two or three communities are needed to capture up to $85\%$ of the user online time spent at the most popular locations. Such a simple spatial structure yields simple theoretical expressions. However, if one wants the model to capture more details (e.g., for detailed simulation), the user can instantiate as many communities as needed to explicitly represent the less visited locations.


\noindent $\bullet$ (1.2 Location of communities) If the map of the target environment is available, one should observe the map and identify the points of attraction in the given environment to assign the communities accordingly. The methods described in \cite{Dart-INFOCOM} could be applied to help choosing the ``hot spots'' on campus, by adding up the time users spend at each location on a 2-D map and identifying the peaks. Alternatively, if the map is not available, one can instantiate communities at {\it random locations}\footnote{Concerning matching with the two mobility properties shown in Fig. \ref{pref-and-reappear}, the actual locations of the communities do not make a difference.}. One way to do so is to simply divide the simulation area into equal-sized grid cells, and assign randomly chosen cells as communities.


\noindent $\bullet$ (1.3 Time period structure) From the curves in Fig. \ref{pref-and-reappear}(b), one observes the {\it re-appearance} periodicity and decides on the time period structure accordingly. Typically, human activities are bounded by daily and weekly schedules so a time period structure shown in Fig. \ref{weekly-schedule} would suffice for most applications. If capturing finer behavior based on time-of-day is necessary, one could additionally split the day into time periods with different mobile node behavior. We illustrate this in our third case study, the human encounter trace.


\noindent \textbf{\underline{STEP 2: Assign Other Parameters}}
After the space/time structure is determined, one has to determine the remaining parameters for each community and time period. This includes $\pi_j^t$, $\overline{D_j^t}$, and  $\overline{L_j^t}$, which represent the stationary  probability (which is calculated after selecting proper $p_{i,j}^t$'s that lead to a desired stationary distribution using simple Markov chain theory), average pause time, and average epoch length, respectively, at community $j$ during time period $t$. These parameters can be determined by referring to the curves in Fig. \ref{pref-and-reappear}. We give some general rules of how the parameters change the curves in Fig. \ref{pref-and-reappear} below. The detailed adjustments we make for each specific case studies will be discussed later.

\noindent $\bullet$ The average epoch length in each community, $\overline{L_j^t}$, should be at least in the same order as the edge length of the community, $C_j^t$. This is to ensure that the end point of the epoch becomes almost independent of its starting point, since the mixing time of the corresponding process becomes quite small. (The motivation for this requirement is to keep the theoretical analysis tractable.)

\noindent $\bullet$ The average duration the node stays in community $j$ is given by $\pi_j^t (\overline{D_j^t} + \overline{L_j^t}/\overline{v})$. The ratio between the durations the node stays in each community shapes the location visiting preference curve in Fig. \ref{pref-and-reappear}(a).

\noindent $\bullet$ The highest peak of the re-appearance probability curve (on the $7$-th day under the weekly schedule) in Fig. \ref{pref-and-reappear}(b) is determined by the weighted average probability of the node appearing in the same community during the same type of time period. This value is $\sum_{t=1}^V \frac{T^t}{\sum_{k=1}^V T^k} \sum_{j=1}^{S^t} (P_j^t)^2$, where $P_j^t$ denotes the fraction of time the node spends in community $j$.

\noindent{\textbf{\underline{STEP 3: Adjust User On-off Pattern (Optional)}}}
The mobility trace generated by the TVC model is an ``always-on'' mobility trajectory (i.e., the mobile nodes are always present somewhere in the simulation field). However, in some situations some nodes might be absent occasionally. For example, in a WLAN setting, nodes (e.g. laptops) are often turned off when travelling from one location to another and the ``off'' time is often not negligible \cite{individual-study}. Thus one may need to make {\it optional} adjustments to turn nodes off in the generated trace, depending on the actual environment to match with. To address this we assign a probability $P_{on,j}$ as the probability for the node to be ``on'' in community $j$. In two of the case studies we present (WLAN and vehicular trace), we utilize this feature as the nodes are not always-on in the actual traces.

Note that it is possible to automate part of the above community and parameter selection. This can be done by feeding the curves in Fig. \ref{pref-and-reappear} and the desired level of matching to a program that executes the above steps. Automatic generation of proper synthetic traces is a direction of our future work.

Next, we look into three specific case studies and apply the fore-mentioned procedure in each case, to display that the TVC model successfully produces synthetic mobility traces with matching characteristics observed in the real traces.

\subsection{WLAN Traces} \label{match_WLAN}

\begin{figure}


\centering
\includegraphics[height=2.2in, angle=-90]{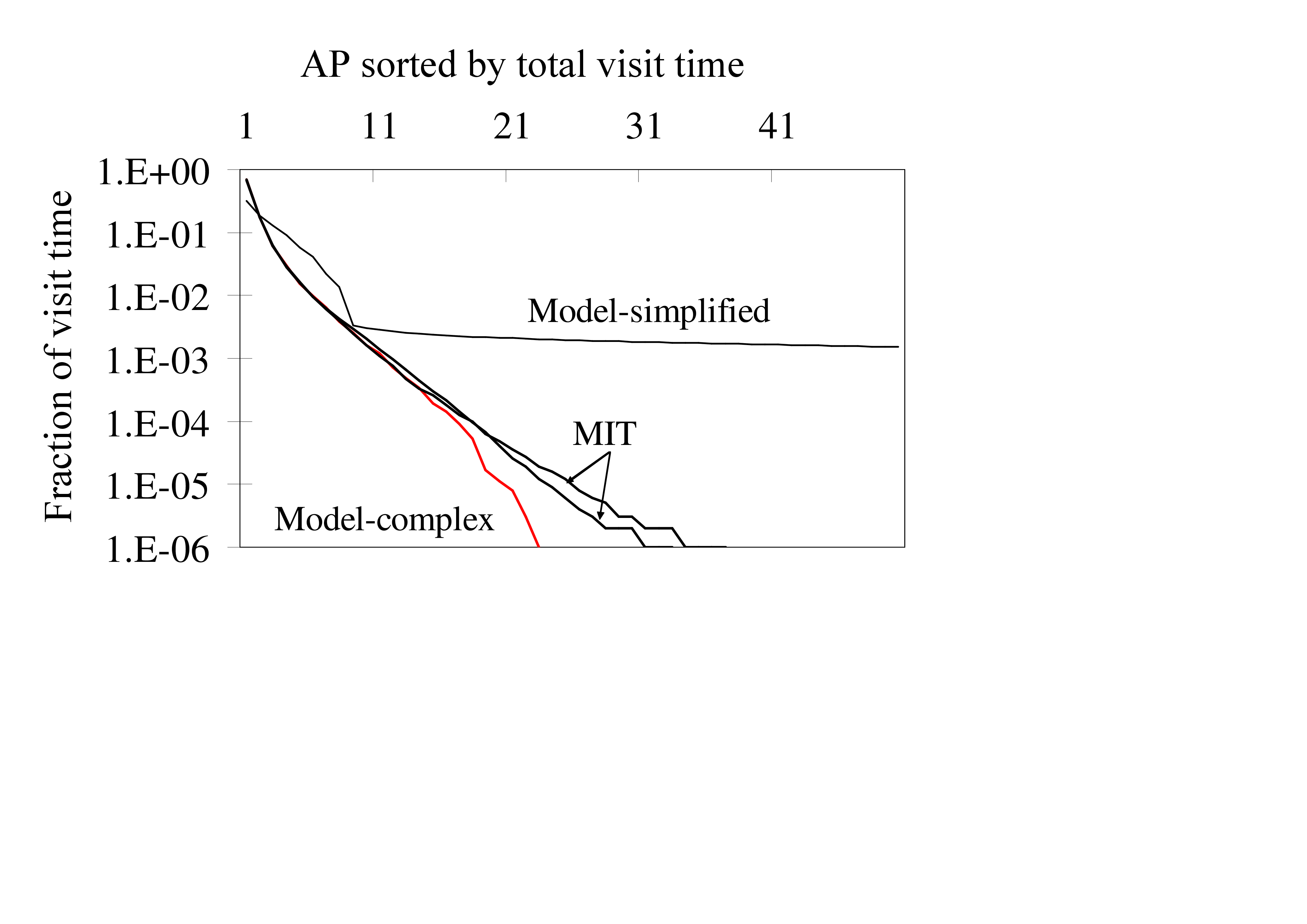}

\vspace{0.05in}

\footnotesize{(a) Skewed location visiting preferences.}


\includegraphics[height=2.2in, angle=-90]{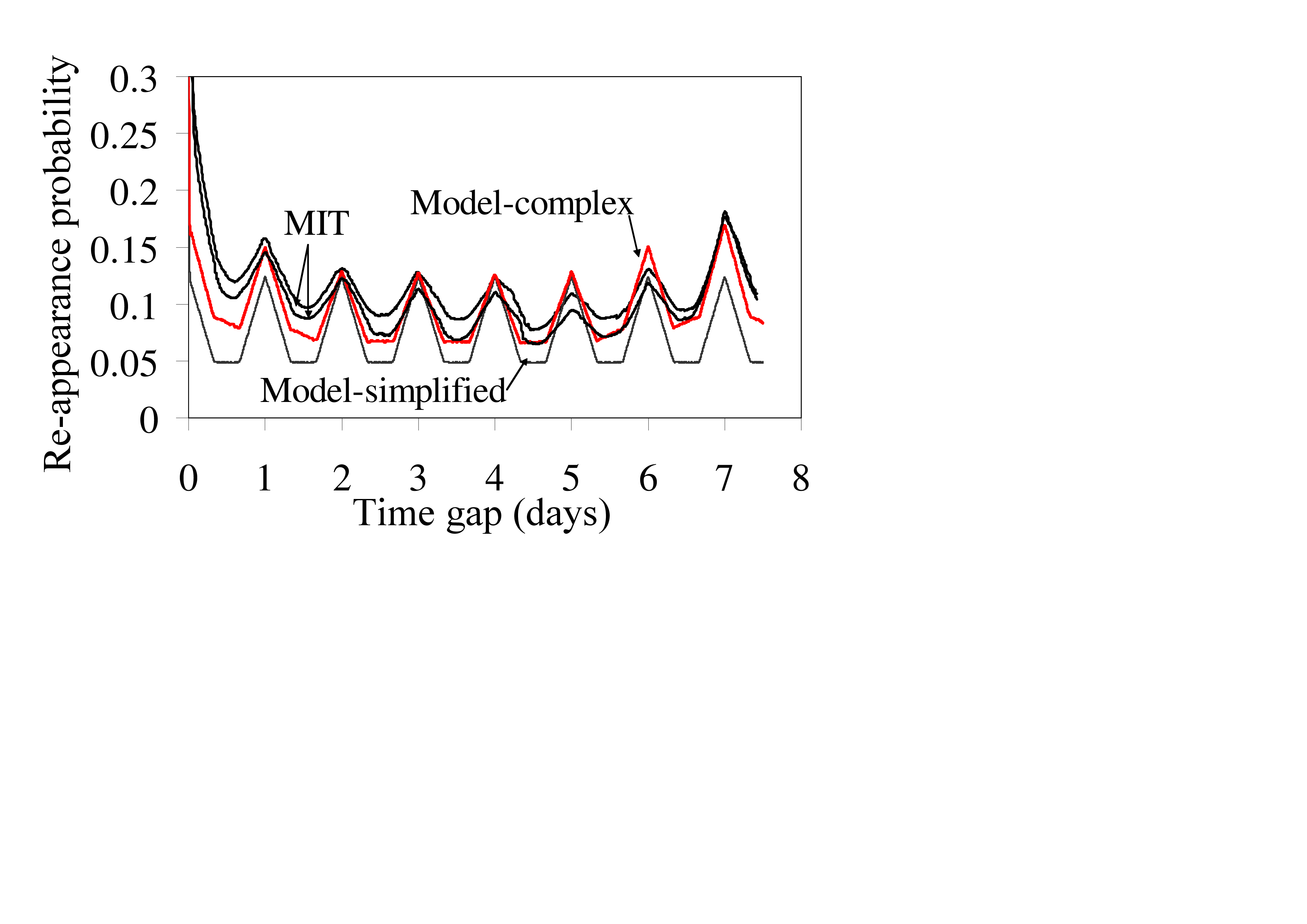}

\vspace{0.05in}

\footnotesize{(b) Periodical re-appearance at the same location.}


\caption{Matching mobility characteristics of the synthetic traces to the MIT WLAN trace.}
\label{model_vs_trace}

\end{figure}

In the first example, we show that the TVC model can re-create the {\it location preferences} and {\it re-appearance probability} curves observed in WLANs. We use the MIT WLAN trace (first presented in \cite{MIT-trace}) as the main example here\footnote{We also achieve good matching with the USC\cite{individual-study} or the Dartmouth\cite{Dart-trace} traces, but do not show it here due to space limitations.}. We split the MIT trace into two halves and generate a matching synthetic trace with observed mobility characteristics from the first half (the training data set). We then compare our synthetic trace with the mobility characteristics of the second half (the validation data set). Note that, the mobility characteristics are similar across the two halves (shown by the two very close thick black curves in Fig. \ref{model_vs_trace}). We generate two synthetic traces with the TVC model, a {\it simplified} one and a {\it complex} one, to display its flexibility to have different levels of matching to the WLAN trace.

The simplified model (shown by thin black curves) uses only one community and two time periods (for the day time and night time), with parameters listed as {\it Model-1} in Table \ref{sim_parameters}. The simple model captures the major trends but still shows several noticeable differences: (a) the tail in the {\it model-simplified} curve in Fig. \ref{model_vs_trace}(a) is ``flat" as opposed to the exponentially diminishing tail of the {\it MIT} curve. (b) the peaks in the {\it model-simplified} curve in Fig. \ref{model_vs_trace}(b) are of equal heights. 

We can improve the matching between the synthetic trace and the real trace by adding complexity in both space and time, with the following detailed procedure. (STEP1): We divide the simulation area into 10-by-10 grid cells. Since we want to have a close match with the curve in Fig. \ref{model_vs_trace}(a), we assign randomly $15$ of the cells as communities to each node (Intuitively, this number corresponds to the number of distinct access points that a person may connect to on a university campus over a period of one month.). For the time period structure we use the simple weekly structure shown in Fig. \ref{weekly-schedule}, allocating $8$ hours for day time (TP1, TP3) and $16$ hours for night time (TP2), as this trace is collected from a corporate environment. (STEP2 and STEP3): In the actual WLAN trace the nodes are ``on'' only for a low percentage of time. We capture this phenomenon with an additional parameter, $P_{on,j}^t$, the probability the node is ``on'' in state $j$. In WLAN, the nodes are typically ``on'' (i.e., appear at the current location) when they are not moving. Under this on-off pattern,  $P_{on,j}^t =  \overline{D_j^t}/(\overline{D_j^t} + \overline{L_j^t}/\overline{v})$. We then consider the on-off pattern and parameter assignment jointly. (1) We first assign the same $\overline{D_j^t}, \overline{L_j^t}, P_{on,j}^t$ to all communities, then assign $\pi_j^t$ with  a value equal to the fraction of time spent at the $j$-th location in Fig. \ref{pref-and-reappear}(a). This assignment strategy makes the node ``on'' for the same amount of time in each community during each visit, and the total time in each community (and hence the observed location visiting preference curve) is therefore determined by the value of $\pi_j^t$. (2) Due to the on-off pattern, the peak value in the re-appearance probability curve becomes $\sum_{t=1}^V \frac{T^t}{\sum_{k=1}^V T^k}  \sum_{j=1}^{S^t} (P_j^t)^2 (P_{on,j}^t)^2$.  To shape the re-appearance probabilities, we adjust the $\overline{D_j^t}$ values, which, in turn, adjust the values of $P_{on,j}^t$ and set the re-appearance probabilities to the desirable values to match with the curve in Fig. \ref{pref-and-reappear}(b). Note that by adjusting the $\overline{D_j^t}$ values in a consistent manner among all communities we do not change the location visiting probability curve that has already been matched in the previous step.

As it is evident from Fig. 4, this model, which is labeled Model-complex and corresponds to the red curves in the plot, yields synthetic traces whose characteristics match very closely with those of the MIT trace.

\subsection{Vehicle Mobility Traces} \label{match_VANET}

\begin{figure}

\begin{minipage}[t]{1.7in}

\centering
\includegraphics[height=1.7in,angle=-90]{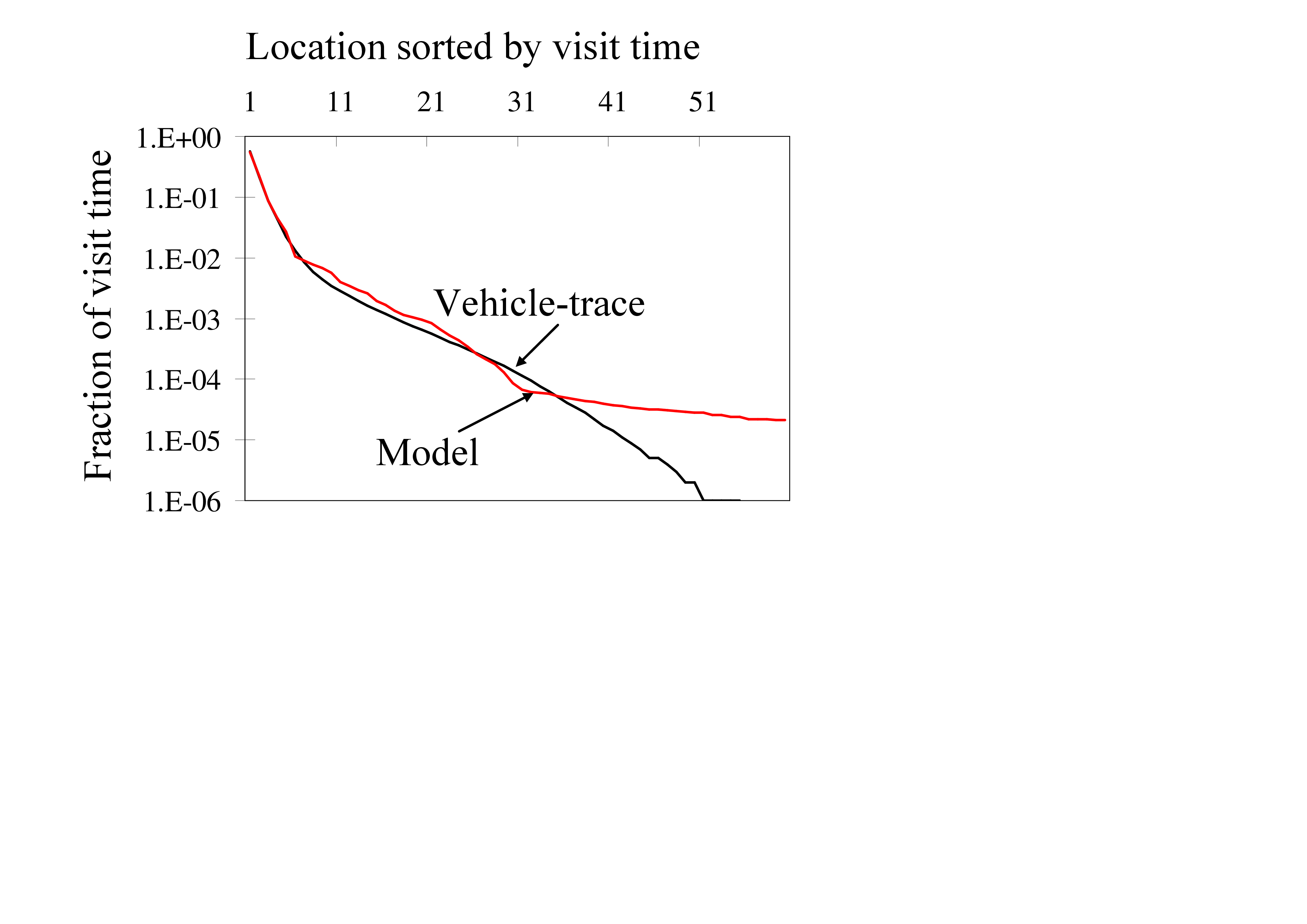}

\vspace{0.05in}

\footnotesize{(a) Location visiting preferences.}

\end{minipage}
\hfill
\begin{minipage}[t]{1.7in}

\centering
\includegraphics[height=1.7in,angle=-90]{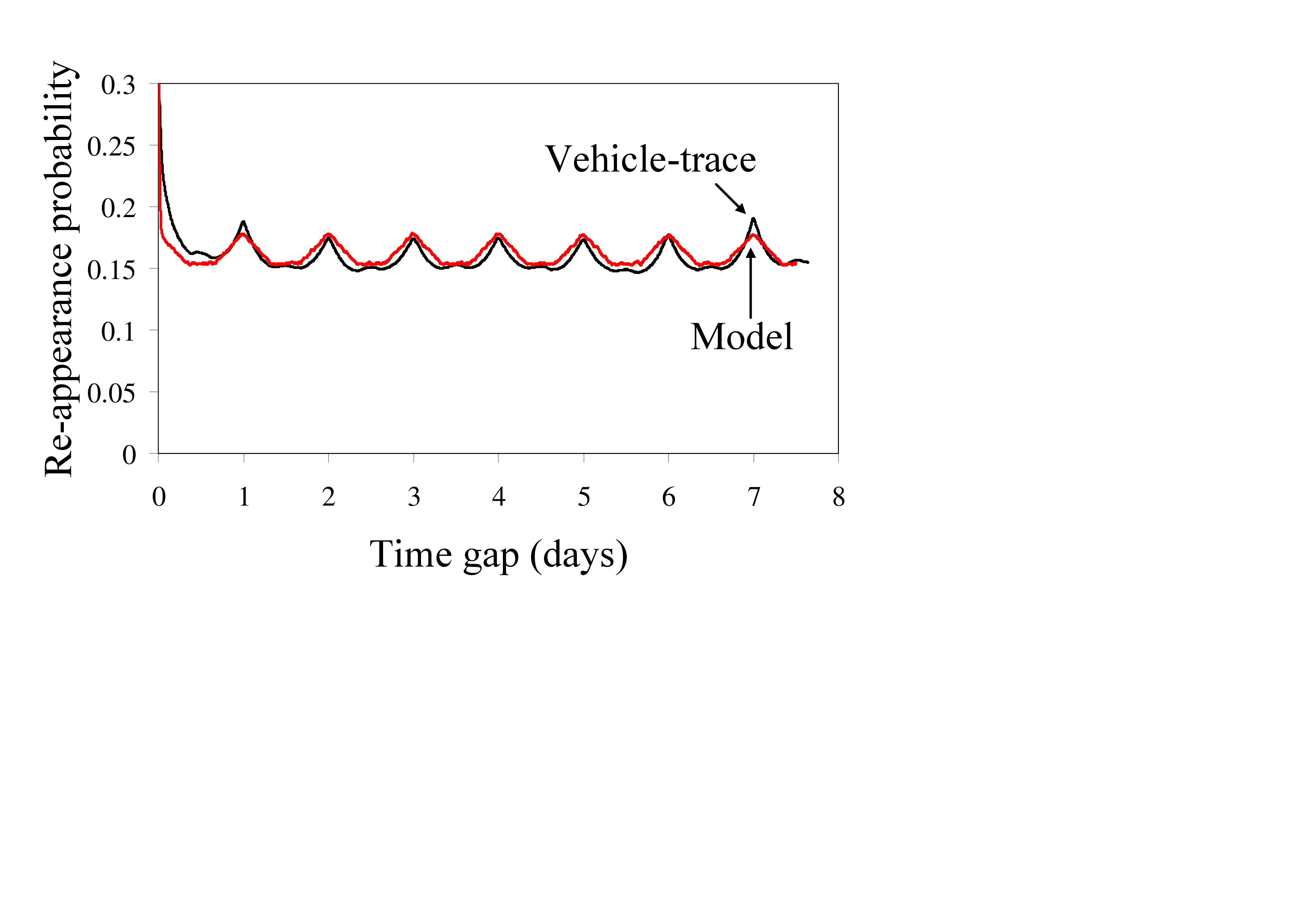}

\vspace{0.05in}

\footnotesize{(b) Periodical re-appearance.}

\end{minipage}
\hfill

\caption{Matching mobility characteristics of the synthetic trace to the vehicle mobility trace.}
\label{model_vs_taxi}

\end{figure}

In this example we display that {\it skewed location visiting preferences} and {\it periodical re-appearance} are also prominent mobility properties in vehicle mobility traces. We obtain a vehicle movement trace from \cite{cab-spotting}, a website that tracks participating taxis in the greater San Francisco area. We process a 40-day trace obtained between Sep. 22, 2006 and Nov. 1, 2006 for 549 taxis to obtain their mobility characteristics. The results are shown in Fig. \ref{model_vs_taxi} with the label {\it Vehicle-trace}. It is interesting to observe that the trend of vehicular movements is very similar to that of WLAN users in terms of these two properties.


We use $30$ communities and the weekly time schedule in (STEP1). We need more communities for this trace as the taxis are more mobile and visit more places than people on university campuses. From the actual trace, we discover that the taxis are offline (i.e., not reporting their locations) when not in operation. Hence we assume that the nodes are ``on'' only when they are moving. The pause times between epochs are considered as breaks in taxi operation. Therefore in (STEP3),  $P_{on,j}^t = (\overline{L_j^t}/\overline{v})/(\overline{D_j^t} + \overline{L_j^t}/\overline{v})$, and we adjust the parameters in a similar way as described in the previous section. The curves in Fig. \ref{model_vs_taxi} with label {\it Model} match with the curves with {\it Vehicle-trace} label well. As a final note, although vehicular movements are generally constrained by streets and our TVC model does not capture such microscopic behaviors, designated paths and other constraints could still be added in the model's map (for vehicular or human mobility) without losing its basic properties. We defer this for future work.

\subsection{Human Encounter Traces} \label{match_enc_trace}

\begin{figure}

\begin{minipage}[t]{1.7in}

\centering
\includegraphics[height=1.7in,angle=-90]{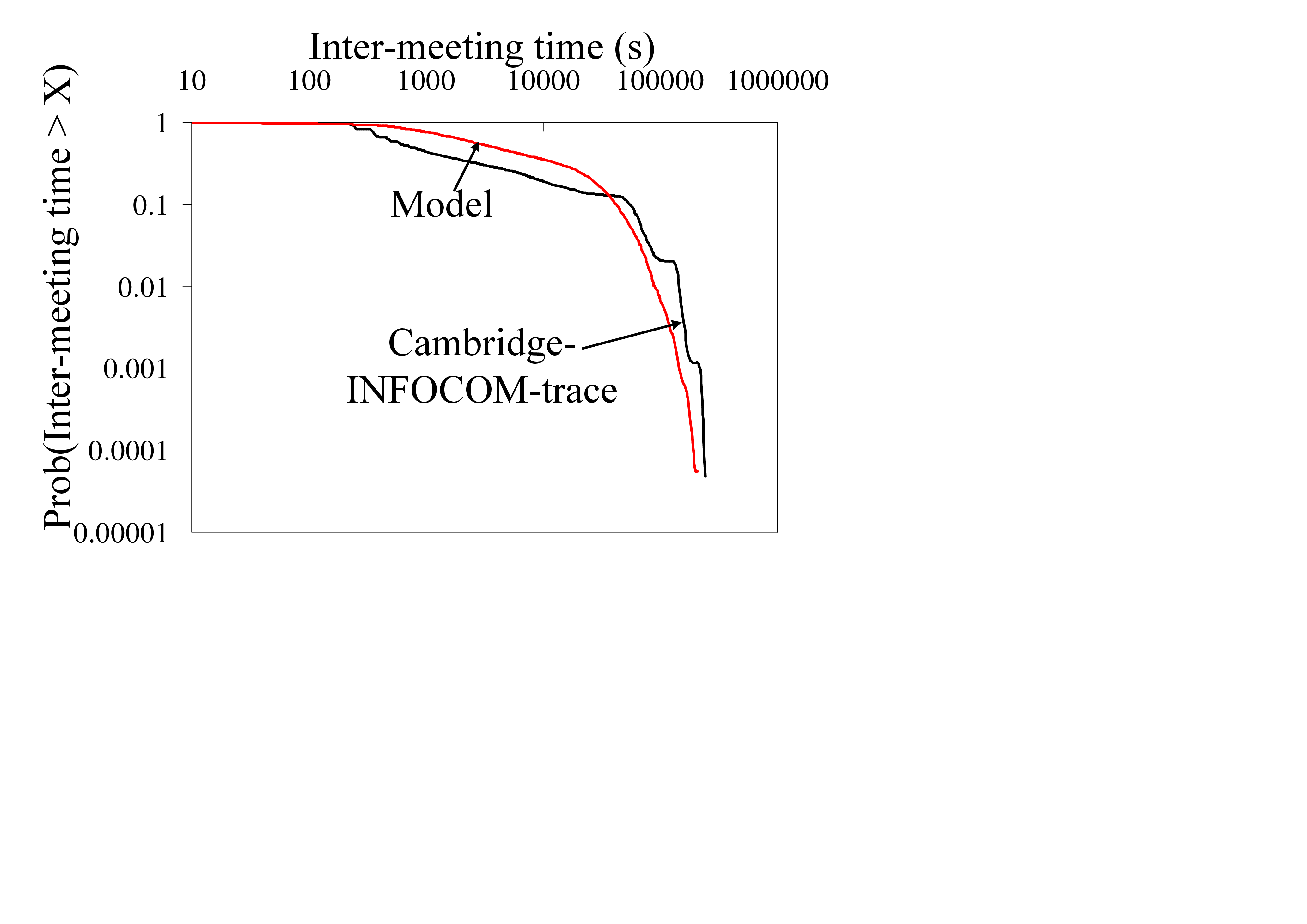}

\footnotesize{(a) Inter-meeting Time.}

\end{minipage}
\hfill
\begin{minipage}[t]{1.7in}

\centering
\includegraphics[height=1.7in,angle=-90]{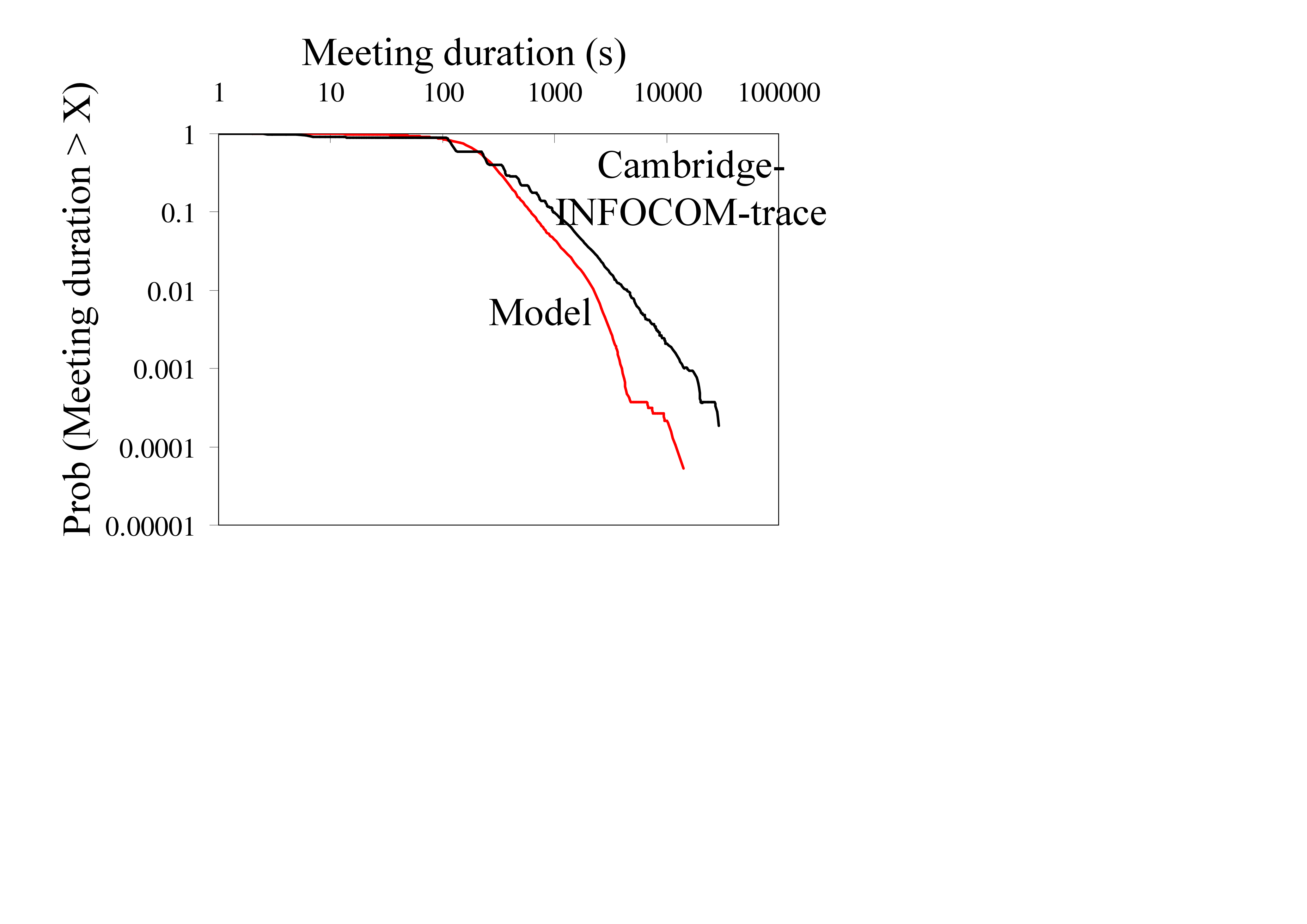}

\footnotesize{(b) Encounter duration.}

\end{minipage}
\hfill

\caption{Matching {\it inter-meeting time} and {\it encounter duration} distributions with the encounter trace.}
\label{meeting_dur}
\end{figure}

In this example, we show that the TVC model is generic enough to mimic the encounter properties of mobile human networks observed in an experiment performed at INFOCOM 2005~\cite{Cambridge-trace}. In this experiment, wireless devices were distributed to $41$ participants at the conference to log encounters between nodes (i.e., coming within Bluetooth communication range) as they moved around the premises of the conference area. The inter-meeting time and the encounter duration distributions of all $820$ pairs of users obtained from this trace are shown in Fig. \ref{meeting_dur} with label {\it Cambridge-INFOCOM-trace}. 

To mimic such behaviors using our TVC model, we observe the conference schedule at INFOCOM, and set up a daily recurrent schedule with five different types of time periods (STEP1): technical sessions, coffee breaks, breakfast/lunch time, evening, and late night (see \cite{simu_code_download} for the detailed parameters). For each time period we set up communities as the conference rooms, the dining room, etc. We also generate a community that is far away from the rest of the communities for each node and make the node sometimes isolated in this community to capture the behavior of patrons skipping part of the conference. In (STEP2), we use the theory presented in section \ref{theory} to adjust the parameters and shape the inter-meeting time and encounter duration curves. For example, a stronger tendency for nodes to choose roaming epochs (setting larger $\pi_r^t$) would increase the meeting probability (see, e.g., Eq. (\ref{unit-time-meeting-prob_full})), hence reducing inter-meeting times. Finally, since the devices used to collect the encounter traces are always-on, we do not apply any changes to the synthetic trace (STEP3). We randomly generate $820$ pairs of users and show their corresponding distributions of the inter-meeting time and the encounter duration in Fig. \ref{meeting_dur} with label {\it Model}.  It is clear that our TVC model has the capability to reproduce the observed distributions, even if it is not constructed explicitly to do so. This displays its success in capturing the decisive factors of typical human mobility.


It is clear from the cases studied here that the TVC model is flexible to capture mobility characteristics from various environments well. In addition, with the respective configuration, it is possible to generate synthetic traces with much larger scale (i.e., more nodes) than the empirical ones while maintaining the same mobility characteristics. It is also possible to generate multiple instances of the synthetic traces with the same mobility characteristics to complement the original, empirically collected trace.





\section{Theoretical Analysis of the TVC Model} \label{theory}

So far, we have established the flexibility of the TVC
model in terms of its ability to reproduce the properties observed
in qualitatively different mobility traces.  Yet, one of the biggest
advantages of our model is that, in addition to the realism, it is
also analytically tractable with respect to some important
quantities which determine protocol performance. In the rest of this
paper, we focus on demonstrating this last point.


We start here by deriving the theoretic expressions of various
properties of the proposed mobility model assuming the nodes are
always ``on''. The properties of interest are defined below.


\noindent $\bullet$ The {\it average node degree} is the average number of nodes residing within the communication range of a given node. This is a quantity of interest due to its implication on the success rate of various tasks (e.g. geographic routing~\cite{geographic-degree}) in mobile ad hoc networks.

\noindent $\bullet$ The {\it hitting time} is the time it takes a node, starting from the stationary distribution, to move within transmission range of a fixed, randomly chosen target coordinate in the simulation field.

\noindent $\bullet$ The {\it meeting time} is the time until two mobile nodes, both starting from the stationary distribution, move into the transmission range of each other. The {\it hitting and meeting times} are of interest due to their close relationship to the performance of DTN routing protocols.



We note that a preliminary version of some of the theoretical
derivations presented here appear under a special case of our TVC
model in~\cite{conf-version-TVC} (that model included one community
and two time periods only). Here, we generalize all derivations for
any community and time-period structure. We start with a useful
lemma that calculates the probability of a node to reside in a
particular state.


\begin{lem}\label{move-prob}
{\it The probability that a node moves, pauses (after the completion
of an epoch) in state $j$, or performs a transitional epoch at any
given time instant during time period $t$, respectively, is:

\begin{equation} \label{PM}
P_{move,j}^t = \pi_j^t  (\overline{L_{j}^t}/\overline{v_{j}^t}) / \Psi,
\end{equation}

\vspace{-0.1in}

\begin{equation} \label{PP}
P_{pause,j}^t = \pi_j^t  \overline{D_j^t} { } /  { } \Psi,
\end{equation}

\vspace{-0.1in}

\begin{equation} \label{Ptx}
P_{tr}^t = \sum_{ k = 1}^{S^t} \pi_k^t \sum_{\forall n} p_{k,n}^t \overline{L_{tr(k,n)}}/\overline{v_{k}^t}  { } /  { }  \Psi.
\end{equation}

where $\Psi = \sum_{ k = 1}^{S^t}  \pi_k^t (\overline{L_{k}^t}/\overline{v_{k}^t} + \overline{D_k^t} + \sum_{\forall n} p_{k,n}^t \overline{L_{tr(k,n)}}/\overline{v_{k}^t} )$ and $\overline{L_{tr(k,n)}}$ is the average length of a transitional epoch from community $k$ to community $n$.
}

\end{lem}

\begin{proof}
The probability for a node to be in state $j$ ($\pi_j^t$) can be
easily derived with Markov chain theory from the state transition
probabilities ($p_{i,j}^t$). The above result follows from the ratio
of the average durations of the moving part
($\overline{L_{j}^t}/\overline{v_{j}^t}$) and the pause part
($\overline{D_j^t}$) of regular epochs, and the transitional epochs
($\overline{L_{tr(k,n)}}/\overline{v_{k}^t}$), weighted by the
probabilities of the states. The expected length of the transitional
epochs, $\overline{L_{tr(k,n)}}$, can be calculated as follows. Note
that if community $n$ contains community $k$, no transitional epoch
is needed (i.e., $L_{tr(k,n)}=0$). The transitional epoch is thus
needed for a roaming node to go back to a smaller community, and as
the previous roaming epoch ends at a random location in the whole
simulation field, by symmetry, the expected length of the
transitional epoch is the average length to move to the center of
the simulation field from a random point in the simulation field.
Numerical analysis concludes $\overline{L_{tr}} = 0.3826N$ in this
case.
\end{proof}

Note that the above stationary probabilities can be calculated for
each time period and node separately. We use $P_j^t (i)$ to denote
the probability that node $i$ is in state $j$ during time period $t$
(i.e., $P_j^t (i) = P_{move,j}^t (i) + P_{pause,j}^t (i)$).

\subsection{Derivation of the Average Node Degree} \label{AND}

The average node degree of a node is defined as the expected number
of nodes falling within its communication range. Each node
contributes to the average node degree independently, as nodes make
independent movement decisions.
\begin{lem} \label{node_in_comm_range_prob}
{\it Consider a pair of nodes, $a$ and $b$. Assume further that, in
time period $t$, community $j$ of node $a$ and community $k$ of node
$b$ overlap with each other for an area $A(a_j^t, b_k^t)$. Then, the
contribution of node $b$ to the average node degree of node $a$,
when $a$ resides in its $j$-th community and $b$ resides in its
$k$-th community, is given by
\begin{align}
\label{node_degree_one_to_one}
\begin{split}
 \frac{\pi K^2}{{C_j^t}^2 (a)} \frac{A(a_j^t, b_k^t)}{{C_k^t}^2 (b)},
\end{split}
\end{align}
where $K$ is the communication range of the nodes.}
\end{lem}

\begin{proof}
Since nodes follow random direction movement in each epoch,
they are uniformly distributed within each community (i.e., they 
are at any point within the community equally likely). The
probability for node $b$ to fall in the $j$-th community of node $a$
is simply the ratio of the overlapped area over the size of the
$k$-th community of node $b$. Node $a$ covers any given point in its
community equal-likely, hence given node $b$ is in the overlapped
area, it is within the communication range of node $a$ with
probability $\pi K^2/{C_j^t}^2 (a)$.
\end{proof}



Following the same principle in Lemma \ref{node_in_comm_range_prob}, we include all community pairs and arrive at the following Theorem.
\begin{thm} \label{node_degree}
{\it The average node degree of a given node $a$ is
\begin{align} \label{node_degree_all}
\begin{split}
\sum_{\forall Comm_j^t (a)}  P_j^t (a) \sum_{\forall b}  \sum_{ \forall Comm_k^t (b) }  P_j^t (b) \frac{\pi K^2}{{C_j^t}^2 (a)} \frac{A(a_j^t, b_k^t)}{{C_k^t}^2 (b)}.
\end{split}
\end{align}}
\end{thm}
\begin{proof}
Eq. (\ref{node_degree_all}) is simply a weighted average of the node
degree of node $a$ conditioning on its states. For each state with
probability $P_j^t (a)$, the expected node degree is a sum over all
other nodes' probability of being within the communication range of
node $a$, again conditioning on all possible states. Transitional
epochs are treated the same way as roaming epochs here. That
is, when considering a node in the transitional state with
probability $P_{tr}^t$, it has equivalent contribution to the node
degree as when it is in the roaming state (i.e., the
node appears uniformly in the simulation field during transitional epochs since the it moves
from anywhere in the simulation field back to the local community.).
 Hence, with probability
$P_{tr}^t + P_{roam}^t$\footnote{If the node has no roaming state in
this time period, then we consider only $P_{tr}^t$.}, the node has
an effective community size of the simulation field, $N$.
\end{proof}

\begin{cor} \label{node_degree_uniform}
{\it In the special case when all nodes choose their communities uniformly at random among the simulation field, Eq. (\ref{node_degree_all}) degenerates to $\sum_{\forall b} \frac{\pi K^2}{N^2}$.}
\end{cor}
\begin{proof}
This result follows from the fact that a randomly chosen community is anywhere in the simulation field equally likely.

\end{proof}

\subsection{Derivation of the Hitting Time} \label{HT}

In the calculation of the average
node degree, the dependence between consecutive epochs did not
affect the derivation. In fact, only the stationary occupancy
probabilities $\pi_{j}^t$ and $P_{j}^{t}$ (i.e. the probability of being
found in community $j$) are needed, since we were looking only at a
random \emph{snapshot} of the model. In the case of hitting and
meeting times, we are
interested in counting the number of epochs until a given target
coordinate is found (``hit''). Our approach is to try to calculate
the ``hit probability'' for a given epoch, and then count the number
of such epochs needed on average until the destination point is hit.
If these probabilities were independent, then one could use a simple
geometric distribution to derive the result. However, (1) consecutive
epochs are strongly related, as the ending point of one epoch is,
naturally, the beginning point of the next. This introduces a
\emph{seeming} dependency between the hit probabilities of
consecutive epochs, complicating the derivation. What is more, (2) the
\emph{transition} between communities (and epochs performed in each)
are governed by the TVC model's Markov chain and the respective
community transition probabilities $p_{i,j}^{t}$. Thus, looking only at
the stationary probabilities for ``choosing'' the next community $j$
(as in the previous section) no longer suffices. Finally,
(3) the \emph{transitional epochs} themselves introduce further
complications, as they cannot, in this case, be handled as regular
in-community or even roaming epochs.

The above three observations introduce dependencies that, at first
glance, complicate our task. Nevertheless, we will show how these
dependencies can be ``washed out'' under a (minimally restricting)
set of assumptions, and that stationary probabilities still suffice
to derive a simple formula for the respective hitting time that
holds \emph{in the limit}. The basis of our argument is found in the
proof of Lemma~\ref{hitting probability}, upon which the rest of
results in this section depend (In a nutshell, the fast mixing of the mobility process takes care of  (1), the large number of epochs required to hit a target takes care of (2) in the limit, and the dominance of local and roaming epochs over transitional epochs takes care of (3).). In Section~\ref{match_theory_sim},
we show that the accuracy of our theory is not compromised by these
assumptions and that our derivations introduce little error in most
practical scenarios considered.

The sketch of the derivation of the hitting time is as follows: (i)
We first condition on the relative location of the target coordinate
with respect to a node's communities (Lemma~\ref{overall-HT}). We
identify all possible sub-cases (i.e. whether the target is inside or
outside one or more of the node's communities). A
target inside a community is, naturally, expected to be found faster
than a target outside all communities. Using simple
geometric arguments, we calculate the probability of each of these
sub-cases (Lemma~\ref{ht-case-probability}) 
and take the weighted average of all sub-cases and
the respective hitting time (to be calculated per sub-case). (ii) For a given sub-case,
we derive the expected number of epochs (and the expected number of
time units) until the target is found (Lemma~\ref{hitting
probability}). (iii) Finally, we introduce the time-period factor, and
account for the total number of time
periods needed to hit the target
(Theorem~\ref{conditional-ht_full_thm}).

The most influential factor for the hitting time is whether the
target coordinate is chosen inside the node's communities. We denote
the possible relationships between the target location and the set
up of communities during time period $t$ as the set $\Omega^t$. Note
that the cardinality of set $\Omega^t$ is at most $2^{S^t}$ (i.e.
for each of the $S^t$ communities, the target coordinate is either
in or out of it). 


\begin{lem} \label{overall-HT}
{\it By the law of total probability, the average hitting time can be written as
\begin{equation} \label{HT_overall_full}
HT = \sum_{w^1 \in \Omega^1, ..., w^V \in \Omega^V} P(w^1, ..., w^V) HT(w^1, ..., w^V),
\end{equation}
\noindent where $w^1, w^2, ..., w^V$ denote one particular
relationship (i.e. a combination of $\{out, in\}^{S^t}$) between the
target coordinate and the community set up during time period $1,2,
...,V$, respectively. Functions $P(\cdot)$ and $HT(\cdot)$ denote
the corresponding probability for this scenario and the conditional
hitting time under this scenario, respectively. Note that each
sub-case $\lbrace w^1, w^2, ..., w^V\rbrace$ is disjoint from all
other sub-cases.}
\end{lem}

To evaluate Eq. (\ref{HT_overall_full}), we need to calculate $P(w^1, ..., w^V)$ and $HT(w^1, ..., w^V)$ for each possible sub-case $(w^1, ..., w^V)$.

\begin{lem} \label{ht-case-probability}
{\it If the target coordinate is chosen independent of the communities and the communities in each time period are chosen independently from other periods, then
\begin{equation}
P(w^1, ..., w^V) = \Pi_{t=1}^{V} P(w^t),
\end{equation}
\noindent where $ P(w^t) = A(w^t) / N^2$, i.e., the probability of a
sub-case $w^t$ is proportional to the area $A(w^t)$ that corresponds
to the specific scenario $w^t$, which is a series of conditions of
the following type: ($\lbrace target \in comm^t_1 \rbrace, \lbrace
target \notin comm^t_2 \rbrace, ..., \lbrace target \in comm^t_S
\rbrace$).}
\end{lem}

\begin{proof}
The result follows from simple geometric arguments.
\end{proof}

The first step for calculating $HT(w^1, ..., w^V)$ is to derive the
unit-time hitting probability in time period $t$ under target
coordinate-community relationship $w^t$, denoted as $P_h^t(w^t)$.

\begin{lem} \label{hitting probability}
{\it For a given time period $t$ and a specific scenario $w^t$,
\begin{equation}
\label{unit-time-hitting-prob_full}
P_{h}^t (w^t)= \sum_{j=1}^{S^t} I(target \in Comm_j^t | w^t) P_{move,j}^t 2K \overline{v_{j}^t}/{C_{j}^t}^2,
\end{equation}
\noindent where $I(\cdot)$ is the indicator function.
}
\end{lem}

\begin{proof}
In order to calculate the expected hitting time, let us first count
the total number of epochs needed. Let us assume that $N_{e}$ epochs
are needed in total, and let us denote as epoch $E_{k_{m}}(m)$ the
$m$-th epoch in sequence (that is occurring in community
$k_{m}$). Let further, $P(k_{1},k_{2},\dots,k_{m})$ denote the
probability of the specific sequence of epochs occurring. Then, the
probability that the target has not been found after $n$ epochs is

\small
\begin{eqnarray}
P(N_{e} > n) & = & \sum_{k_{1},k_{2},\dots,k_{n}}
P(k_{1},k_{2},\dots,k_{n}) \cdot P(E_{k_{1}}(1) = \mbox{miss}, \nonumber \\
& & E_{k_{2}}(2) = \mbox{miss},\dots,E_{k_{n}}(n) = \mbox{miss)}.
\label{eq:pr1}
\end{eqnarray}
\normalsize

In order to simplify the above equation, we need to deal with the
inherent dependencies introduced by the transition of epochs. First,
since node movement is continuous, the end of one epoch $E_{j}(m)$,
performed in community $j$, is the beginning of the next,
$E_{j}(m+1)$, if performed in the same community \footnote{For the
moment, we will ignore transitional epochs, and assume that all
epochs are performed inside some community; we deal with
transitional epochs later.}. Nevertheless, as explained in
Section~\ref{model-spec}, the expected ``length'' of an epoch
$\overline{L_{j}^t}$ performed in community $j$ is in the order of the
square root of the community size $C_{j}^t$. This is sufficient for
the node to ``mix'' in the community after just one
epoch~\cite{Akis-MOBIHOC}. Consequently, we can write

\small
\begin{eqnarray}
P(N_{e} > n) =  \sum_{k_{1},k_{2},\dots,k_{n}} P(k_{1},k_{2},\dots,k_{n}) \cdot  \prod_{i=1}^{n} P(E_{k_{i}} = \mbox{miss}).
\label{eq:pr2}
\end{eqnarray}
\normalsize

An additional dependency arises from the transitions between
communities and the calculation of term
$P(k_{1},k_{2},\dots,k_{n})$. If epoch $E_{k_{m}}(m)$ is performed
in community $k_{m}$ the next epoch $E_{k_{m+1}}(m+1)$ will be
performed in community $k_{m+1}$ with probability $p_{k_{m},k_{m+1}}$
(the transition probability in the Markov Chain governing the
community transitions in the TVC model). Let us assume that $N_{e}$
denotes again the total epochs needed (of any type) to hit the
target. Further, let there be $l_{j}$ epochs of type $j$ (i.e.
performed in community $j$) in the above mix of $N_{e}$ total
epochs. When $N_{e} \rightarrow \infty$, then $l_{i} \rightarrow
\pi_{i} N_{e}$, that is, the total number of epochs in community $i$
depends {\it only} on the stationary probability of community $i$, $\pi_{i}$.
Thus,

\begin{equation} \label{eq:pr3}
P(k_{1},k_{2},\dots,k_{n}) = \pi_{k_{1}} \cdot \pi_{k_{2}} \dots
\pi_{k_{n}}.
\end{equation}

Consequently, Eq.(\ref{eq:pr2}) becomes

\begin{equation}
P(N_{e} > n) =  \prod_{i=1}^{n} \pi_{k_{i}} \cdot P(E_{k_{i}} =
 \mbox{miss}).
\end{equation}

This implies that, \emph{in the limit}\footnote{In practice, the
requirement is that a large number of epochs is needed on average
until the target is hit. In the sparse networks we're interested in,
this is a reasonable assumption, and as we shall show in
Section~\ref{match_theory_sim} the achieved accuracy is indeed
high.}, the total number of epochs needed to hit the target can be
approximated by a geometric distribution, where the ``average''
epoch has a hit probability of

\begin{equation} \label{eq:pr5}
\sum_{i}^{S^{t}} \pi_{i} \cdot P(E_{i} =  \mbox{hit})
\end{equation}

As the final step, we need to calculate the probability of a given
epoch in community $j$ to hit the target.  Instead of using this per
epoch hit probability, we revert now to what we call the
\emph{unit-step} hit probabilities, $P_h$. The unit-step probability
is the probability of encountering the target exactly within the
next time-unit (rather than within the duration of a whole epoch).
This discrete approximation provides an equivalent formulation to
the above continuous case (see~\cite{Akis-MOBIHOC}), however it is
more convenient to manipulate in the case of time-period boundaries
and meeting times calculated later. (Note that this approximation is
again only possible when the average epoch length is in the order of
the respective community size, ensuring mixing after one epoch.)

Note that the hitting event can only occur when the node is
physically moving inside the community where the target is located\footnote{We
neglect the small probability that the target is chosen out of the
community but close to it, and make the contributions from epochs in
state $j$ zero if the chosen target coordinate is not in community
$j$.}. Whether the target is located inside community $j$ is denoted
using the indicator function $I(target \in Comm_j^t | w^t)$. If the
target is outside the community, then this probability of hit is
zero. If the target lies within community $j$, then when a node
moves with average speed $\overline{v_j}$, on average it covers a
new area of $2K\overline{v_j}$ in unit time. Since a node following
random direction movements visits the area it moves about with equal
probability, and the target coordinate is chosen at random, it falls
in this newly covered area with probability
$2K\overline{v_j}/{C_{j}^t}^2$~\cite{Akis-MOBIHOC}. Hence the
contribution to the unit-time hitting probability by movements made
in state $j$ is $P_{move,j}^t 2K \overline{v_{j}^t}/{C_{j}^t}^2$.
Thus, in Eq.(\ref{eq:pr5}), $\pi_{j}$ is replaced by $I(target \in
Comm_j^t | w^t) P_{move,j}^t$ in the unit-step case, and $P(E_{i} =
hit)$ by $2K \overline{v_{j}^t}/{C_{j}^t}^2$.

As a final remark, the contribution of transitional epochs to the
unit-time hitting probability is not equivalent to other epochs (due
to the dependency of end-points on local communities, which
introduces bias after communities have been chosen). Nevertheless,
in a normal mobility scenario, we expect a node to spend the
majority of its time within one of the communities rather than in
transitional epochs. Specifically, we will assume that community
transition probabilities exhibit a strong \emph{positive
correlation}, that is, if a node resides in community $j$, it has a
higher probability of staying within this community for the next
epoch, rather than leaving. In this case, the total contribution of
transitional epochs is small, and can be safely ignored in order to
not complicate our analysis. The above is a reasonable assumption
for many target scenarios we can imagine; simulation results show further that 
the time a node resides in transitional state
is indeed less than $10\%$ in the scenarios considered, not
significantly affecting the accuracy of the above expression.

\end{proof}

Given the fore-mentioned
assumptions about unit-step hitting probabilities, the corollary below follows.

\begin{cor} \label{period hitting prob}
{\it The probability for at least one hitting event to occur during time period $t$ under scenario $w^t$ is
\begin{equation} \label{time-period-hitting-prob_full}
P_{H}^t (w^t) = 1- (1 - P_{h}^t (w^t) )^{T^t}.
\end{equation}
}
\end{cor}

Finally, using the law of total probability, we derive the conditional hitting time under a specific target-community relationship, $HT(w^1, ..., w^V)$.

\begin{thm} \label{conditional-ht_full_thm}
{\it
\begin{align} \label{ht-total-prob_full}
\begin{split}
HT(w^1, ..., w^V) & = \sum_{t=1}^{V} HT(w^1, ..., w^V | first \; hit \; in \; period \; t ) \cdot \\
& P(w^1, ..., w^V, first \; hit \; in \; period \; t ),
\end{split}
\end{align}

\noindent where the probability for the first hitting event to happen in time period $t$ is
\begin{align} \label{ht-condition-prob_full}
\begin{split}
&P(w^1, ..., w^V, first \; hit \; in \; period \; t ) \\
=& \frac{\Pi_{i=1}^{t-1} (1- P_{H}^i (w^i)) \cdot  P_{H}^t (w^t)}{P},
\end{split}
\end{align}

\noindent and the hitting time under this specific condition is
\begin{align} \label{conditional-ht_full}
\begin{split}
& HT(w^1, ...,w^V | first \; hit \; in \; period \; t ) \\
 = & \sum_{i=1}^{V} T^i \cdot (\frac{1}{P} - 1) + \sum_{i=1}^{t-1} T^i + \frac{1}{P_{h}^t (w^t)},
\end{split}
\end{align}
\noindent where $P = 1 - \Pi_{t=1}^{V} (1- P_{H}^t (w^t))$ is the hitting probability for one full cycle of time periods.}
\end{thm}

\begin{proof}
Eq. (\ref{ht-condition-prob_full}) holds as each cycle of time periods follows the same repetitive structure, and for the first hitting event to occur in time period $t$ it must not occur in time period $1, ..., (t-1)$.
The first term in Eq. (\ref{conditional-ht_full}) corresponds to the expected duration of full time period cycles until the hitting event occurs. Since for each cycle the success probability of hitting the target is $P$, in expectation it takes $1/P$ cycles to hit the target, and there are $1/P-1$ full cycles. The second term in Eq. (\ref{conditional-ht_full}) is the sum of duration of time periods before the time period $t$ in which the hitting event occurs in the last cycle. Finally, the third term is the fraction of the last time period before the hitting event occurs. Note that the last part is an approximation which holds if the time periods we consider are much longer than unit-time.
\end{proof}

\subsection{Derivation of the Meeting Time} \label{MT}

The procedures of the derivation of the meeting time is similar to
that of the hitting time detailed in the last section. In short, we
derive the unit-step (or unit-time) meeting probability, $P_m$, and
the meeting probability for each type of time period, $P_M$, and put
them together to get the overall meeting time in a similar fashion
as in Theorem \ref{conditional-ht_full_thm}.

Similar to Lemma \ref{hitting probability}, we add up the
contributions to the meeting probability from all community pairs
from node $a$ and $b$ in the following Lemma.

\begin{lem}
{\it Let community $j$ of node $a$ and community $k$ of node $b$ overlap with each other for an area $A(a_j^t, b_k^t)$ in time period $t$. Then, the conditional unit-time meeting probability in time period $t$ when node $a$ and $b$ are in its community $j$ and $k$, respectively, is

\begin{align} \label{unit-time-meeting-prob_full}
\begin{split}
& P_{m}^t (a_j^t, b_k^t) = \\
 & P_{move,j}^t (a) P_{move,k}^t (b) \hat{v} \frac{2K \overline{v}}{A(a_j^t, b_k^t)} \frac{A(a_j^t, b_k^t)}{{C_j^t}^2 (a)} \frac{A(a_j^t, b_k^t)}{{C_k^t}^2 (b)}\\
      & + P_{move,j}^t (a) P_{stop,k}^t (b)  \frac{2K \overline{v}}{A(a_j^t, b_k^t)} \frac{A(a_j^t, b_k^t)}{{C_j^t}^2 (a)} \frac{A(a_j^t, b_k^t)}{{C_k^t}^2 (b)} \\
 & + P_{stop,j}^t (a) P_{move,k}^t (b)  \frac{2K \overline{v}}{A(a_j^t, b_k^t)} \frac{A(a_j^t, b_k^t)}{{C_j^t}^2 (a)} \frac{A(a_j^t, b_k^t)}{{C_k^t}^2 (b)}.
\end{split}
\end{align}
}
\end{lem}

\begin{proof}
Equation (\ref{unit-time-meeting-prob_full}) consists of two parts:

(I) Both of the nodes are moving within the overlapped area. This adds the first term in Eq. (\ref{unit-time-meeting-prob_full})
to the meeting probability. The two ratios, $\frac{A(a_j^t, b_k^t)}{{C_j^t}^2 (a)}$ and $\frac{A(a_j^t, b_k^t)}{{C_k^t}^2 (b)}$, capture the probabilities that the nodes are in the overlapped area of the communities. The contribution to the unit-time meeting probability is the product of probabilities of both nodes moving within the overlapped area and the term $\frac{2K \overline{v}}{A(a_j^t, b_k^t)}$, which reflects the covered area in unit time. We use the fact that when both nodes move according to the random direction model, one can calculate the effective (extra) area covered by assuming that one node is static, and the other is moving with the (higher) \emph{relative speed} between the two. This difference is capture with the multiplicative factor $\hat{v}$~\cite{Akis-MOBIHOC}.


(II) One node is moving in the overlapped area, and the other one pauses within the area. This adds the remaining two terms in Eq. (\ref{unit-time-meeting-prob_full})
to the unit-time meeting probability. These terms follow similar rationale as the previous one, with the difference that now only one node is moving. The second term corresponds to the case when node $a$ moves (and $b$ is static), and the third term corresponds to the case when node $b$ moves (and $a$ is static).


The derivation of the unit-time meeting probability between nodes $a$ and $b$ for time period $t$ includes all possible scenarios of community overlap. If node $a$ has $S^t (a)$ communities and node $b$ has $S^t (b)$ communities, there can be at most $S^t (a) S^t (b)$ community-overlapping scenarios in time period $t$. For similar reason detailed in the proof of Lemma \ref{hitting probability}, we neglect the contribution of transitional epochs to the unit-time meeting probability.
\end{proof}

Note that (\ref{unit-time-meeting-prob_full}) is the general form of Equation (13) and (14) in \cite{conf-version-TVC}. If we assume perfect overlap and a single community from both nodes, we arrive at (14). If we assume no overlap, we result in (13). Also note in the general expressions presented in this paper, the whole simulation area is also considered as a community. Therefore we do not have to include a separate term to capture the roaming epochs.

\begin{cor}
{\it The probability for at least one meeting event to occur during time period $t$ is
\begin{align} \label{time-period-meeting-prob_full}
\begin{split}
P_{M}^t  = &1 - \sum_{ \forall (j,k) } \{ P_{ov} ( a_j^t , b_k^t)  \cdot  ( 1 - P_m^t (a_j^t , b_k^t))^{T^t} \},
\end{split}
\end{align}
\noindent where $P_{ov} ( a_j^t, b_k^t)$ is the probability that the community $j$ of node $a$ overlaps with community $k$ of node $b$. This quantity is simply $1$ if the communities have fixed assignments and $A(a_j^t, b_k^t) \neq 0$. If the communities are chosen randomly, this probability can be derived by {\it Lemma 4.5} in \cite{conf-version-TVC}. Due to space constraint, the Lemma is not reproduced here.}
\end{cor}

Finally, similarly to Theorem \ref{conditional-ht_full_thm}, the expected meeting time can be calculated using the results in the Lemmas in this section.

\begin{thm}
{\it The expected meeting time is
\begin{equation} \label{mt-total-prob_full}
MT = \sum_{t=1}^{V} MT(meet \; in \; period \; t ) P(meet \; in \; period \; t ).
\end{equation}
\noindent Where the quantities in the above equation are calculated by
\begin{equation} \label{mt-condition-prob_full}
P(meet \; in \; period \; t ) = \frac{\Pi_{i=1}^{t-1} (1- P_{M}^i) \cdot  P_{M}^t}{Q},
\end{equation}
\begin{equation} \label{conditional-mt_full}
MT(meet \; in \; period \; t )  = \sum_{i=1}^{V} T^i \cdot (\frac{1}{Q} - 1) + \sum_{i=1}^{t-1} T^i + \frac{1}{P_{m}^t},
\end{equation}
where $Q = 1 - \Pi_{i=1}^{V} (1- P_{M}^i)$ is the meeting probability for one full cycle of time periods.
}
\end{thm}
\begin{proof}
The proof is parallel to that of Theorem \ref{conditional-ht_full_thm} and is omitted due to space limitations (see \cite{simu_code_download} for details).

\end{proof}

As a final note, we can easily modify the above theory to account
for potential ``off'' periods (e.g. by introducing a per step or per
epoch ``off'' probability, and a respective multiplicative factor).
Due to space limitations, we do not include here these
modifications.

\begin{table*}
\caption{Parameters for the scenarios in the simulation}
\scriptsize
\label{sim_parameters}

We use the same movement speed for all node: $v_{max} = 15$ and $v_{min} = 5$ in all scenarios. In all cases we use two time periods and they are named as time period $1$ and $2$ for consistency. We only list the parameters for the simple models ($Model \;\;1-4$) here.  Please refer to \cite{simu_code_download} for the details of the generic models ($Model \;\; 5-7$).



\begin{center}

\begin{tabular}{|c||c|c|c|c|c|c|c|c|c|c|c|c|c|c|c|}
\hline
Model name & Description & $N$ & $C_l^1$ & $C_l^2$ & $D_{max,l}$ & $D_{max,r}$ & $\overline{L_{l}}$ & $\overline{L_{r}}$ & $\pi_l^1$ & $\pi_r^1$ & $\pi_l^2$ & $\pi_r^2$ & $T^1$ & $T^2$ \\

\hline
Model $1$ & Match with the MIT trace & $1000$ & $100$ & $100$ & $100$ & $50$ & $80$ & $520$ & $0.714$ & $0.286$ & $0.8$ & $0.2$ & $5760$ & $2880$ \\
\hline
Model $2$ & Highly attractive communities & $1000$ & $200$ & $50$ &  $100$ & $200$ & $52$ & $520$ & $0.667$ & $0.333$ & $0.889$ & $0.111$ & $3000$ & $2000$ \\
\hline
Model $3$ & Not attractive communities & $1000$ & $100$ & $100$ & $50$ & $200$ & $80$ & $800$ & $0.5$ & $0.5$ & $0.667$ & $0.333$ & $2000$ & $1000$ \\
\hline
Model $4$ & Large-size communities & $1000$ & $200$ & $250$ & $50$ & $100$ & $200$ & $800$ & $0.7$ & $0.3$ & $0.889$ & $0.111$ & $2000$ & $1000$ \\
\hline
\end{tabular}

\end{center}
\end{table*}

\section{Validation of the Theory with Simulations} \label{match_theory_sim}

In this section, we compare the theoretical derivations of the previous section against the corresponding simulation results, for various parameter settings. Through extensive simulations with multiple scenarios and parameter settings, we establish the accuracy of the theoretical framework. Due to space limitations, we can only show some examples of the simulation results we have. More complete results can be found at \cite{simu_code_download}.

We summarize the parameters for the tested scenarios in Table \ref{sim_parameters}. We use two different setup of the TVC model for the simulation cases. The parameters listed in Table \ref{sim_parameters} are for the simple models ($Model \;\;1-4$), where we have two time periods with two communities in each time period (one of the communities is the whole simulation field). We also simulate for more generic setup of the TVC model ($Model \;\; 5-7$, refer to \cite{simu_code_download} for its parameters), where we have three communities (one of them is the simulation field) in each time period. For the generic models, we have experimented with two ways of community placement: in a tiered fashion (as drawn in TP2 in Fig. \ref{comm-tp-illustration}), or in a random fashion. Our discrete-time simulator is written in C++.  More details about the simulator, as well as the source code, can be found at \cite{simu_code_download}.

\subsection{The Average Node Degree} \label{AND_sim}

For the average node degree, we create simulation scenarios with $50$ nodes in the simulation area, and calculate the average node degree of each node by taking the time average across snapshots taken every second during the simulation, and then average across all nodes. All the simulation runs last for $60000$ seconds in this subsection.

\begin{figure}
\begin{minipage}[t]{1.7in}

\centering
\includegraphics[height=1.7in, angle=-90]{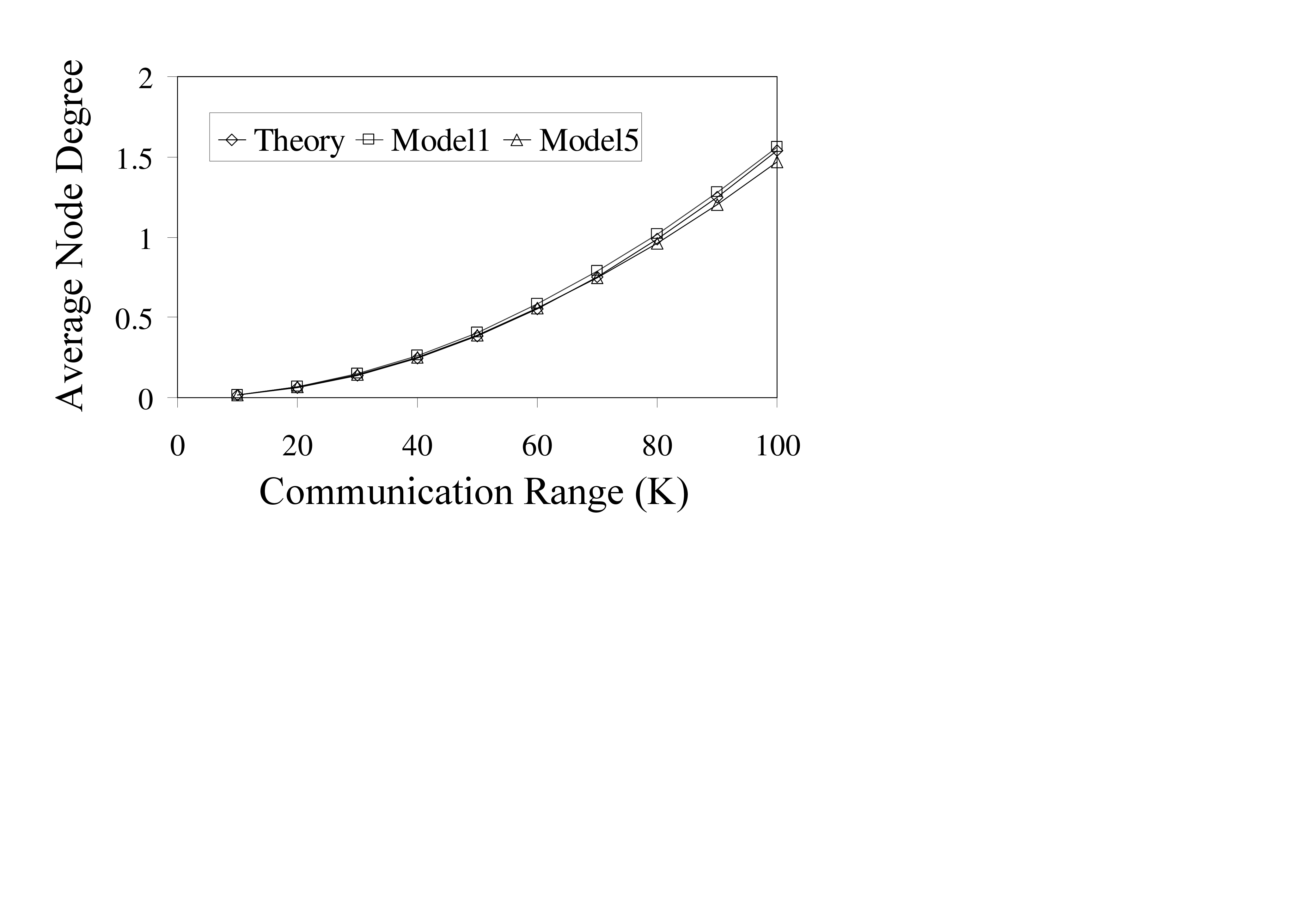}

\vspace{0.05in}

\footnotesize{(a) Randomly placed community.}

\end{minipage}
\hfill
\begin{minipage}[t]{1.7in}

\centering
\includegraphics[height=1.7in, angle=-90]{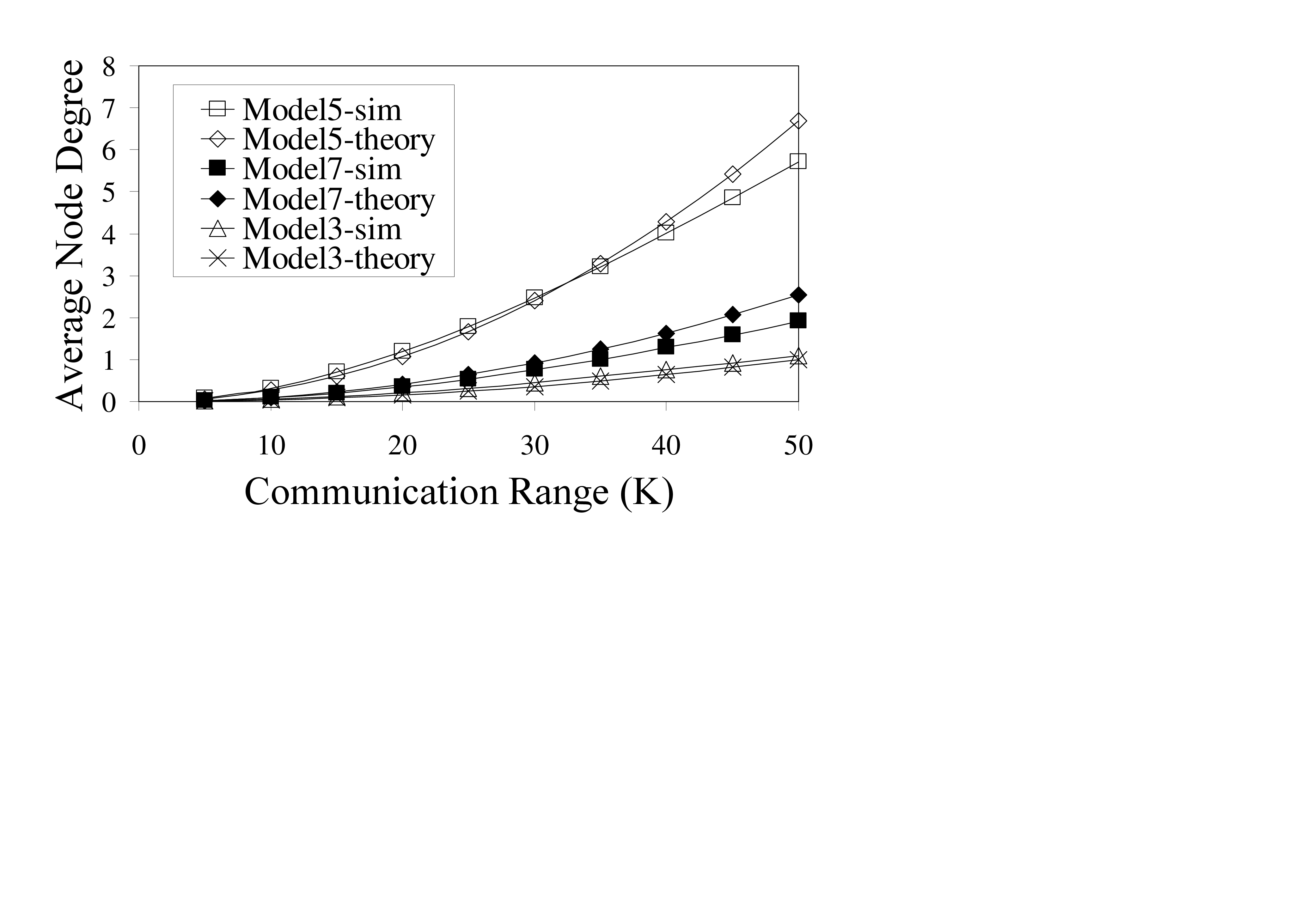}

\vspace{0.05in}

\footnotesize{(b) Fixed communities.}
\end{minipage}
\hfill

\caption{Examples of simulation results (the average node degree).}
\label{node_degree_fig}

\end{figure}

As we show in Corollary \ref{node_degree_uniform}, when the communities are randomly chosen, the average node degree turns out to be the average number of nodes falling in the communication range of a given node, {\it as if all nodes are uniformly distributed}. Hence the average node degree does not depend on the exact choices of community setup (i.e. single, multiple, or multi-tier communities) or other parameters. In Fig. \ref{node_degree_fig} (a), we see the simulation curves follow the prediction of the theory well. 


To make the scenario a bit more realistic, we simulate some more scenarios when the communities are fixed. Among the $50$ nodes, we make $25$ of them pick the community centered at $(300, 300)$ and the other $25$ pick the community centered at $(700, 700)$. We simulate scenarios for all seven sets of parameters, and show some example results in Fig. \ref{node_degree_fig} (b). In the simulations, when the communication ranges are small as compared to the edge of the communities, the relative errors are low, indicating a good match between the theory and the simulation. However, as the communication range increases, the area covered by the communication disk becomes comparable to the size of the community and Eq. (\ref{node_degree_one_to_one}) is no longer accurate since the communication disk extends out of the overlapped area in most cases. That is the reason for the discrepancies between the theory and simulation. Besides {\it Model-3}, we observe at most $20\%$ of error when the communication disk is less than $20\%$ the size of the inner-most community, indicating that our theory is valid when the communication range is relatively small.


\subsection{The Hitting Time and the Meeting Time} \label{HTMT_sim}

We perform simulations for the hitting and the meeting times for $50,000$ independent iterations for each scenario, and compare the average results with the theoretical values derived from the corresponding equations (i.e. (\ref{HT_overall_full}) and (\ref{mt-total-prob_full})). To find out the hitting or the meeting time, we move the nodes in the simulator indefinitely until they hit the target or meet with each other, respectively.

\begin{figure}
\begin{minipage}[t]{1.7in}

\centering
\includegraphics[height=1.7in, angle=-90]{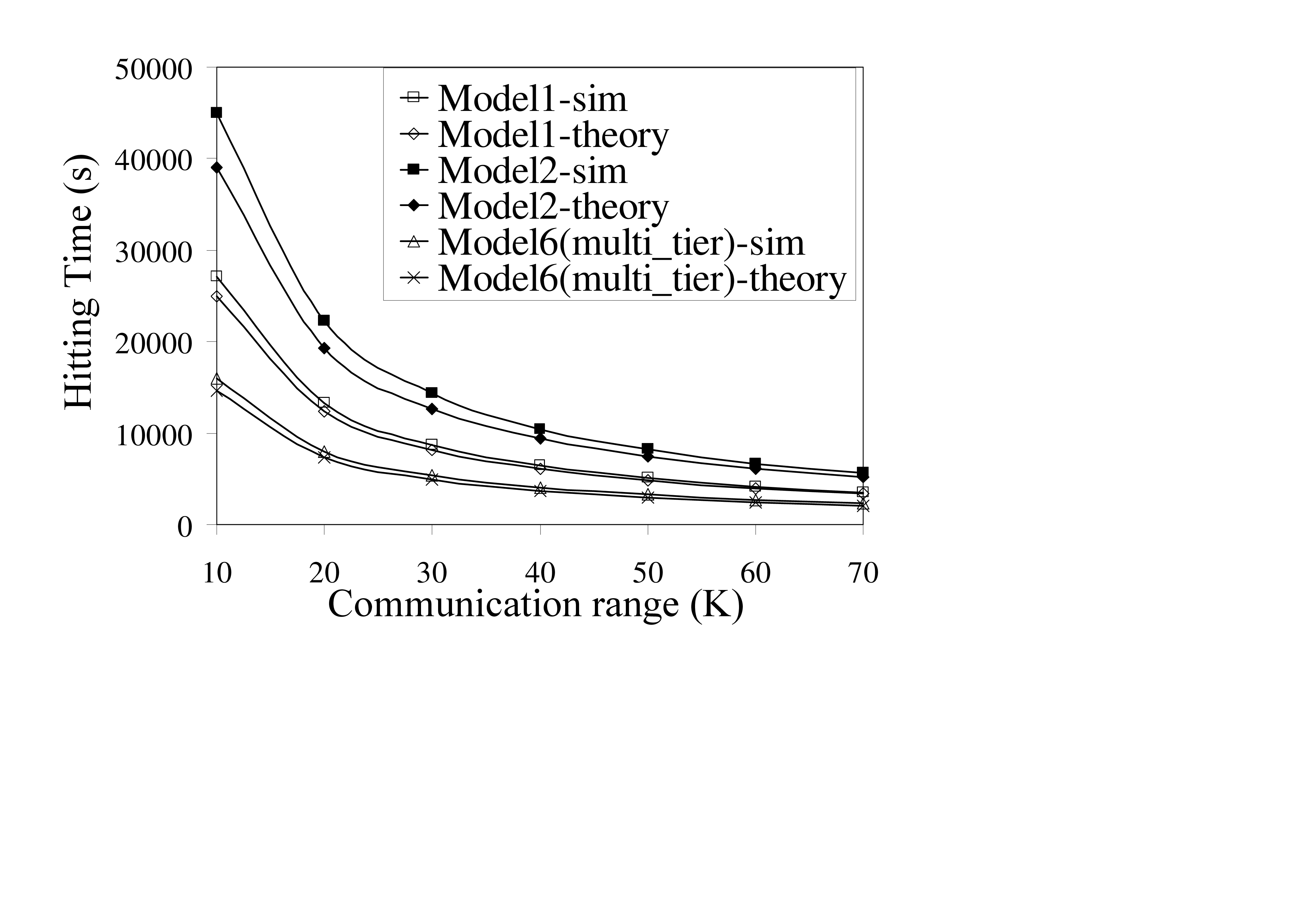}

\footnotesize{(a) Hitting Time.}

\end{minipage}
\hfill
\begin{minipage}[t]{1.7in}

\centering
\includegraphics[height=1.7in, angle=-90]{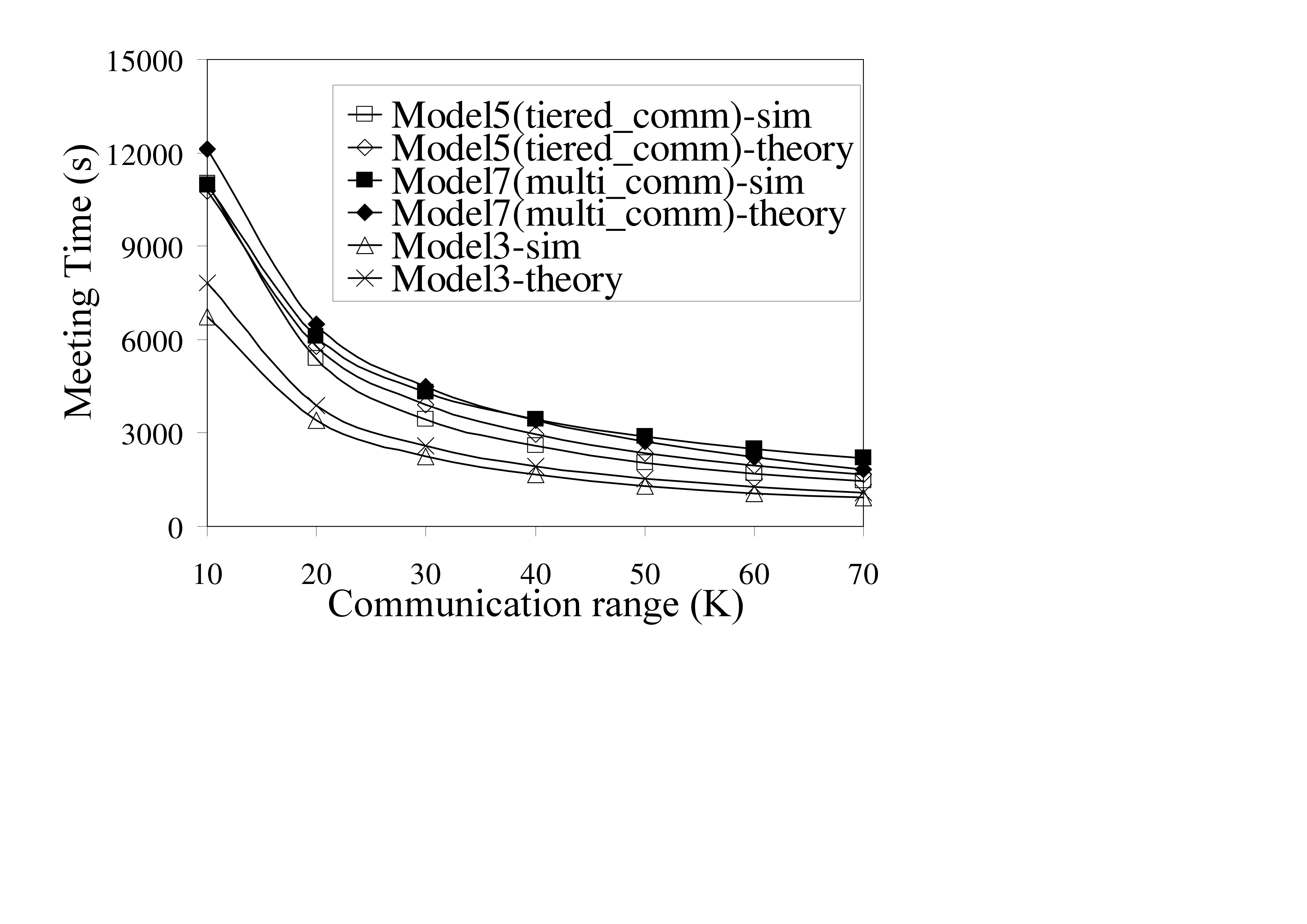}

\footnotesize{(b) Meeting Time.}
\end{minipage}
\hfill

\caption{Examples of simulation results.}
\label{relative_error}

\end{figure}

Again we show some example results in Fig. \ref{relative_error}. For all the scenarios (including the ones not listed here), the relative errors are within acceptable range. The absolute values for the error are within $15\%$ for the hitting time and within $20\%$ for the meeting time. For more than $70\%$ of the tested scenarios, the error is below $10\%$ (refer to \cite{simu_code_download} for other figures). These results display the accuracy of our theory under a wide range of parameter settings. The errors between the theoretical and simulation results are mainly due to some of the approximations we made in the various derivations. For example, the approximation of the hitting and meeting processes with discrete, unit-time Bernoulli trials is valid only for the epochs that are long enough (in the order of community size) and if there are a lot of epochs. Furthermore, there exist some border effects -- when a node is close to the border of a community, it could also ``see'' some other nodes outside of the community if its transmission range is large enough. However, we have chosen to ignore such occurrences to keep our analysis simpler. Nevertheless, as shown in the figures, the errors are always within acceptable ranges, justifying our simplifying assumptions.

\section{Using Theory for Performance Prediction} \label{perf_prediction}

Although the various theoretical quantities derived for the TVC model in Section \ref{theory} are interesting in their own merit, they are particularly useful in predicting protocol performance, which in turn can guide the decisions of system operation. We illustrate this point with two examples in this section.

\subsection{Estimation of the Number of Nodes Needed for Geographic Routing} \label{nodes_needed}

It has been shown in geographic routing that the average node degree determines the success rate of messages delivered~\cite{geographic-degree}. Thus, using the results of Section \ref{AND} we can estimate the number of nodes (as a function of the average node degree) needed to achieve a target performance for geographic routing, for a given scenario.

\begin{figure}
\begin{minipage}[t]{1.7in}

\centering
\includegraphics[height=1.7in, angle=-90]{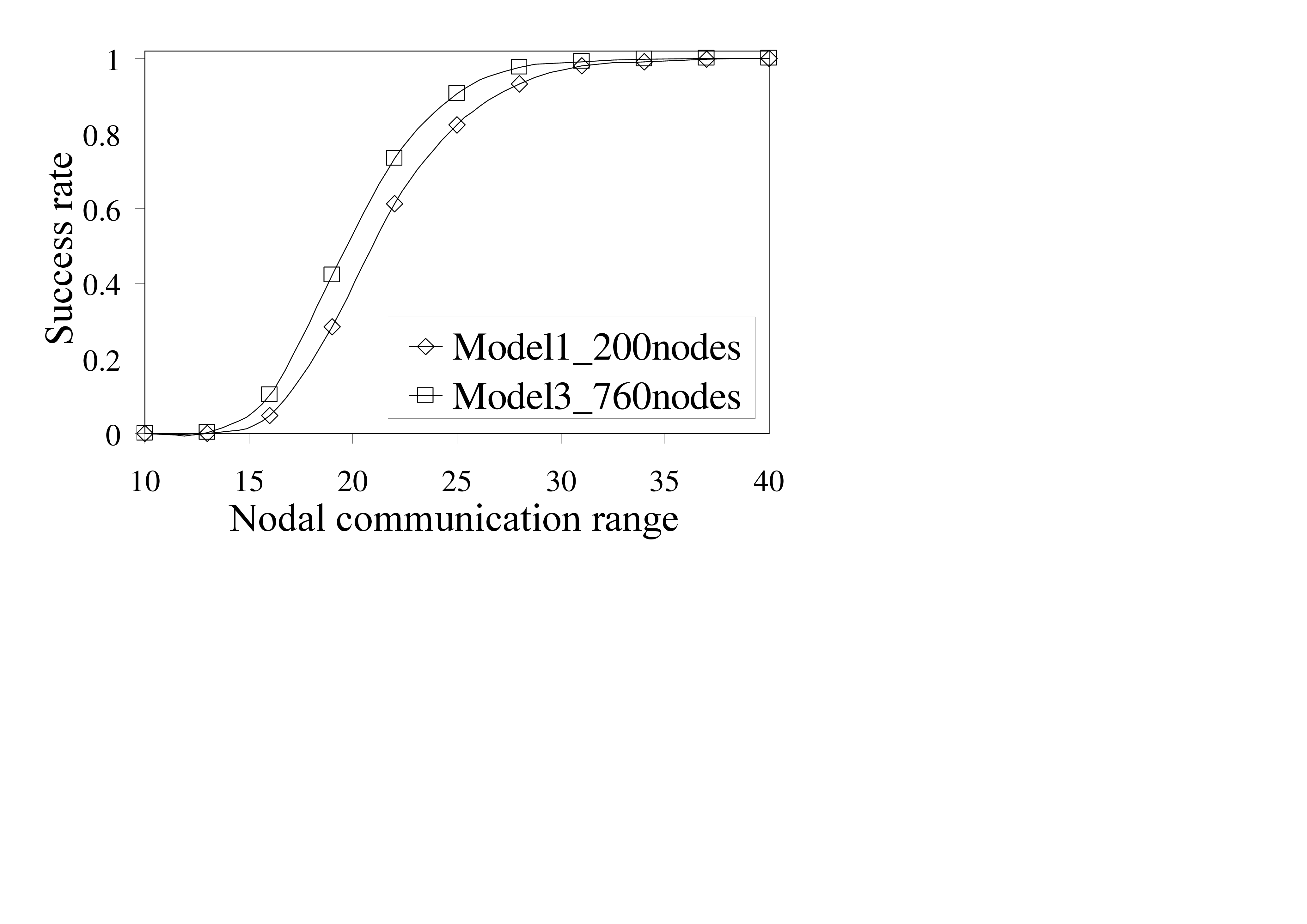}

\caption{Geographic routing success rate under different mobility parameter sets and node numbers.}
\label{geo_routing_success_rate}

\end{minipage}
\hfill
\begin{minipage}[t]{1.7in}

\centering
\includegraphics[height=1.7in, angle=-90]{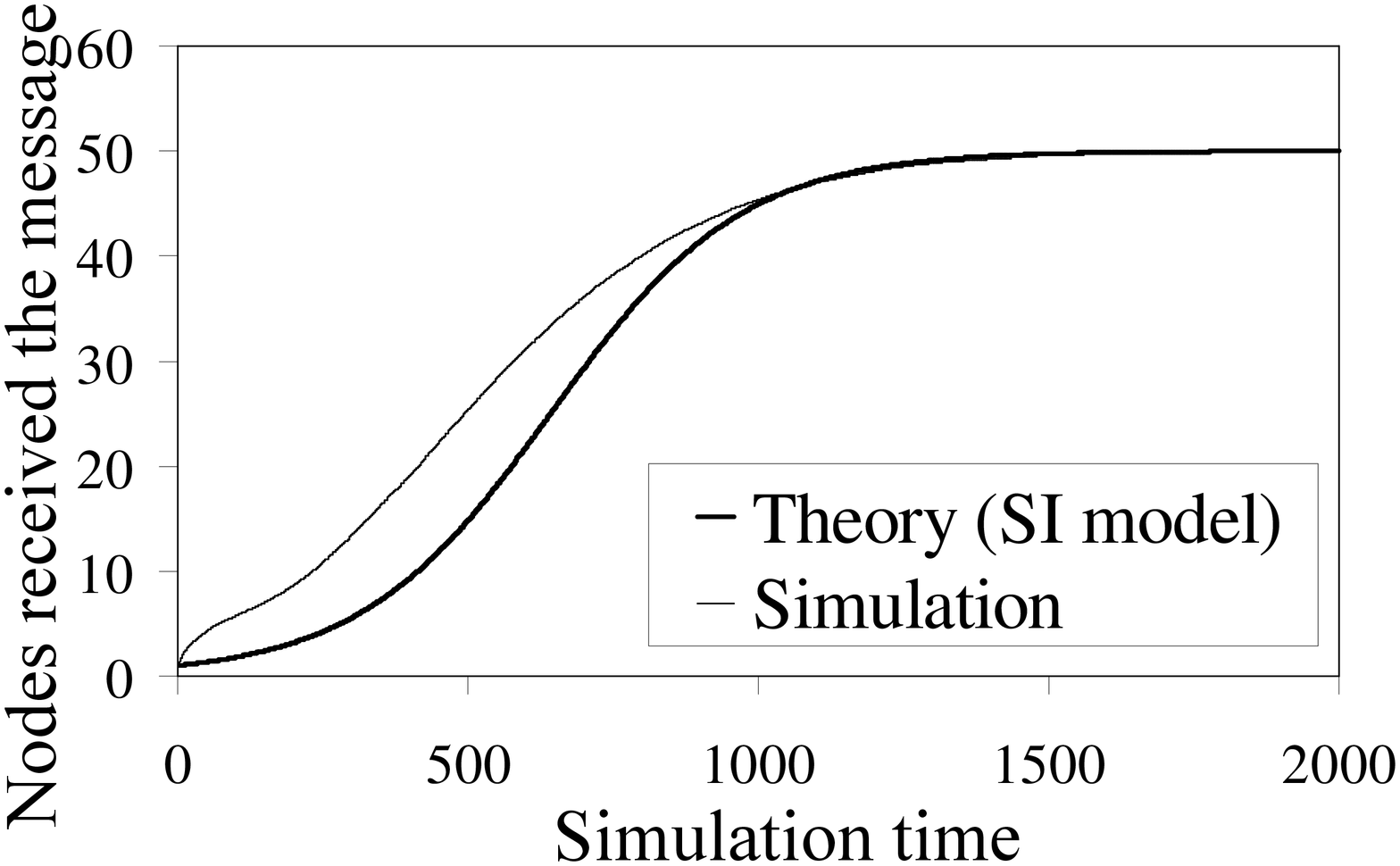}

\caption{Packet propagation with epidemic routing with {\it two} node groups with different communities.}
\label{2comm_epidemic}

\end{minipage}
\hfill

\end{figure}

We consider the same setup as in Section \ref{AND_sim}, where half of the nodes are assigned to a community centered at $(300, 300)$
and the other half are assigned to another community centered at $(700, 700)$. We are interested in routing messages across one of the communities, from coordinate $(250, 250)$ to coordinate $(350, 350)$ with simple geographic routing (i.e., greedy forwarding only, without face routing \cite{GPSR}). Using simulations we obtain the success rate of geographic routing under various communication ranges when $200$ nodes move according to the mobility parameters of {\it Model-1} (Table \ref{sim_parameters}). Results are shown in Fig. \ref{geo_routing_success_rate} (each point is the percentage of success out of $2000$ trials). If we assume the mobility model is different, say {\it Model-3}, we would like to know how many nodes we need to achieve similar performance. Using Eq. (\ref{node_degree_all}) we find that $760$ nodes are needed to create a similar average node degree for {\it Model-3}. To validate this, we also simulate geographic routing for a scenario where $760$ nodes follow {\it Model-3}. Comparing the resulting message delivery ratio for this scenario to the original scenario ($200$ nodes with {\it Model-1}) in Fig. \ref{geo_routing_success_rate}, we see that similar success rates are achieved under the same transmission range, which confirms the accuracy of our analysis.

\subsection{Predicting Message Delivery Delay with Epidemic Routing} \label{SIR_prediction}

Epidemic routing is a simple and popular protocol that has been proposed for networks where nodal connectivity is intermittent (i.e., in  Delay Tolerant Networks)~\cite{epidemic}. It has been shown that message propagation under epidemic routing can be modeled with sufficient accuracy using a simple fluid-based model~\cite{epidemic-perf-ana}. (Note that its performance has also been analyzed using Markov Chain~\cite{inter-meeting} and Random Walk~\cite{single-copy} models.) This fluid model has been borrowed from the Mathematical Biology community, and is usually referred to as the SI (Susceptible-Infected) epidemic model. The gist of the SI model is that the rate by which the number of ``infected'' nodes increases (``infected'' nodes here are nodes who have received a copy of the message) can be approximated by the product of three quantities: the number of already infected nodes, the number of susceptible (not yet infected) nodes, and the pair-wise contact rate, $\beta$ (assuming nodes meet independently -- this contact rate is equivalent to the unit-step meeting probabilities calculated in (\ref{unit-time-meeting-prob_full})). Thus, one could plug-in these meeting probabilities into the SI model equations and calculate the delay for epidemic routing. Yet, in the TVC model (and often in real life) there are multiple groups of nodes with different communities, and thus different pair-wise contact rates that depend on the community setup. For example, nodes with the same or overlapping communities tend to meet much more often than nodes in far away communities. For this reason, we extend the basic SI model to a more general scenario.

We consider the following setup in the case study: We use {\it Model-3} (Table \ref{sim_parameters}) for the mobility parameters. A total of $M=50$ nodes are divided into two groups of $25$ nodes each. One group has its community centered at $(300, 300)$ and the other at $(700, 700)$. One packet starts from a randomly picked source node and spreads to all other nodes in the network. The propagation of the message can be described by the following equations:

\begin{equation} \label{SIR-2comm}
\left \{
\begin{array}{l}
\frac{d I_1(t)}{dt} = \beta_{ov} I_1(t)S_1(t) + \beta_{no\_ov}I_2(t)S_1(t) \\
\frac{d I_2(t)}{dt} = \beta_{ov} I_2(t)S_2(t) + \beta_{no\_ov}I_1(t)S_2(t) \\
S_1(t) + I_1(t) = M/2 \\
S_2(t) + I_2(t) = M/2. \\
\end{array}
\right.
\end{equation}

\noindent where $S_x(t)$ and $I_x(t)$ denote the number of susceptible and infected nodes at time $t$ in group $x$, respectively. Parameters $\beta_{ov}$ and $\beta_{no\_ov}$ represent the pair-wise unit-time meeting probability when the communities are overlapped (i.e., for nodes in the same group) and not overlapped (i.e., nodes in different groups), respectively. We use Eq. (\ref{unit-time-meeting-prob_full}) to obtain these quantities. This model is an extension from the standard SI model~\cite{epidemic-perf-ana} and similar extensions can be made for more than two groups~\cite{sapon-2class}. The first equation governs the change of infected nodes in the first group. Notice that the infection to susceptible nodes in the group ($S_1(t)$) can come from the infected nodes in the same group ($I_1(t)$) or the other group ($I_2(t)$). We can solve the system of equations in (\ref{SIR-2comm}) to get the evolution of the total infected nodes in the network. As can be seen in Fig. \ref{2comm_epidemic}, the {\it theory} curve closely follows the trend in the {\it simulation} curve. This indicates first that scenarios generated by our mobility model are still amenable to fluid model based mathematical analysis (SI), despite the increased complexity introduced by the concept of communities. It also shows that results produced thus can be used by a system designer to predict how fast messages propagate in a given network environment. This might, for example, determine if extra nodes are needed in a wireless content distribution network to speed up message dissemination.

As a final note, in addition to epidemic routing, the theoretical results for the hitting and meeting times could be applied to predict the delay of various other DTN routing protocols (see e.g. \cite{Akis-MOBIHOC,single-copy,epidemic-perf-ana}), for a large range of mobility scenarios that can be captured by the TVC model.




\section{Conclusion and future work} \label{conclusion}

We have proposed a {\it time-variant community mobility model} for wireless mobile networks. Our model preserves common mobility characteristics, namely {\it skewed location visiting preferences} and {\it periodical re-appearance} observed in empirical mobility traces. We have tuned the TVC model to match with the mobility characteristics of various traces (WLAN traces, a vehicle mobility trace, and an encounter trace of moving human beings), displaying its flexibility and generality. A mobility trace generator of our model is available at \cite{simu_code_download}. In addition to providing realistic mobility patterns, the TVC model can be mathematically analyzed to derive several quantities of interest: the {\it average node degree}, the {\it hitting time} and the {\it meeting time}. Through extensive simulations, we have verified the accuracy of our theory.


In the future we would like to further analyze the performance of various routing protocols (e.g., \cite{single-copy, multi-copy}) under the time-variant community mobility model. We also would like to construct a systematic way to automatically generate the configuration files, such that the communities and time periods of nodes are set to capture the inter-node encounter properties we observe in various traces (for example, the Small World encounter patterns observed in WLAN traces~\cite{group-study}).




\begin{IEEEbiography}[{\includegraphics[width=1in,height=1.25in,clip]{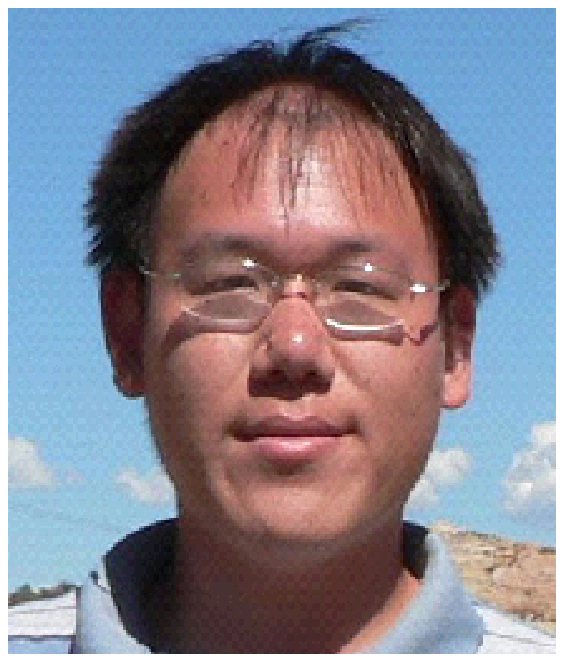}}]{Wei-jen Hsu}
was born in Taipei, Taiwan, in March 1977. He received the B.S. degree in Electrical Engineering and the M.S. degree in Communication Engineering, respectively, from National Taiwan University, in June 1999 and June 2001. He received the Engineer degree in Electrical Engineering from University of Southern California, in August 2006. He is currently a Ph.D. student in the CISE Department, University of Florida. 
His main research interest involves the utilization of realistic measurement data in various tasks in computer networks, including user modeling and behavior-aware protocol design.
\end{IEEEbiography}

\begin{IEEEbiography}[{\includegraphics[width=1in,height=1.25in,clip,keepaspectratio]{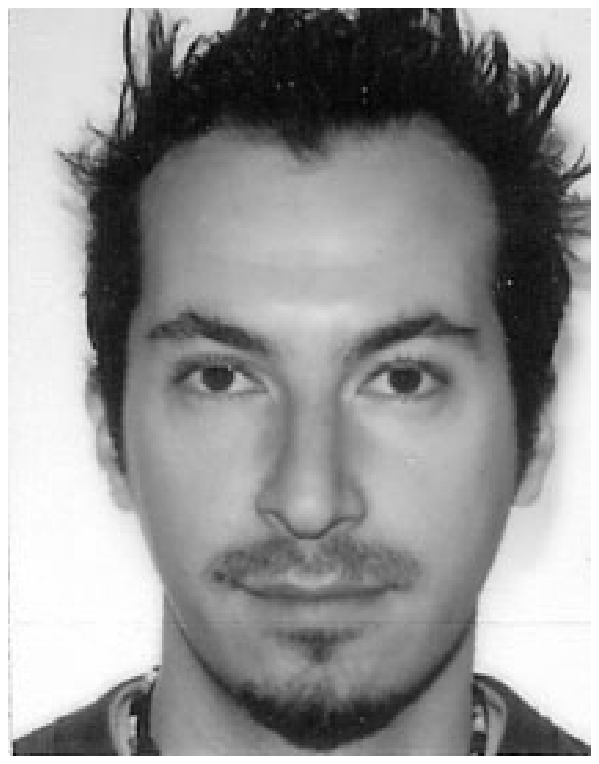}}]{Thrasyvoulos Spyropoulos}
was born in Athens, Greece, in July 1976. He received the Diploma in
Electrical and Computer Engineering from the National Technical
University of Athens, Greece, in February 2000. In May 2006, he
received the Ph.D degree in Electrical Engineering from the
University of Southern California. In 2006-07, he was a post-doctoral
researcher at INRIA, Sophia-Antipolis. He is currently a senior researcher
with the Swiss Federal Institute of Technology (ETH), Zurich.
\end{IEEEbiography}

\begin{IEEEbiography}[{\includegraphics[width=1in,height=1.25in,clip,keepaspectratio]{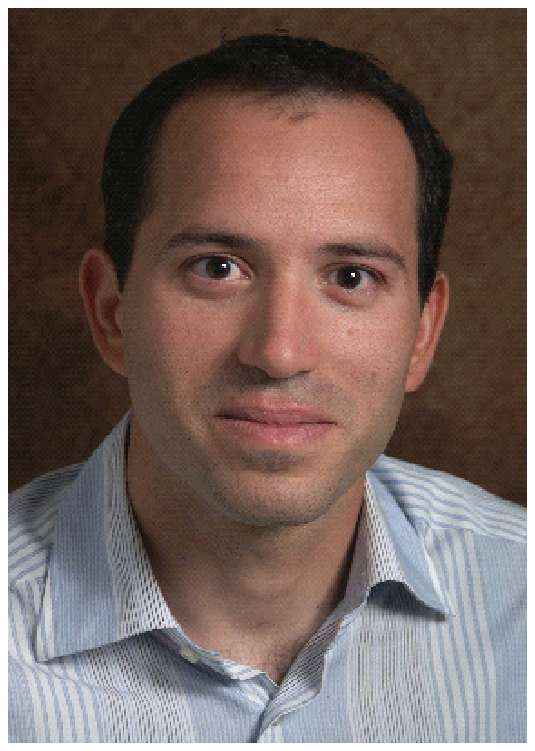}}]{Konstantinos Psounis}
Konstantinos Psounis is an assistant professor of EE and CS at the University of Southern California. He received his 
first degree from National Technical University of Athens, Greece,
in 1997, and the M.S. and Ph.D. degrees from Stanford University in 1999 and 2002 respectively.
Konstantinos models and analyzes the performance of a variety of networks,
and designs methods and algorithms to solve problems related to
such systems. He is the author of more than 40 research papers, 
has received faculty awards from NSF and the Zumberge
foundation, and has been a Stanford graduate fellow throughout his graduate
studies.

\end{IEEEbiography}

\begin{IEEEbiography}[{\includegraphics[width=1in,height=1.25in,clip,keepaspectratio]{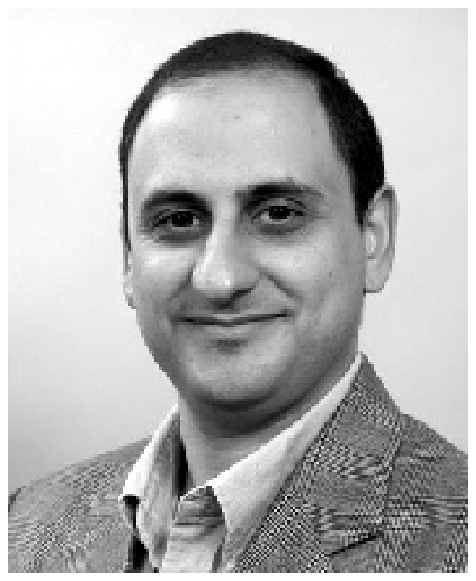}}]{Ahmed Helmy}
Dr. Ahmed Helmy received his Ph.D. in Computer Science (1999), M.S. in Electrical Engineering (EE) (1995) from the University of Southern California (USC). He is Associate Professor and director of the wireless networking lab at the CISE Dept, University of Florida. From 1999 to 2006, he was faculty with EE-USC. He was a key researcher in the network simulator (NS-2) and the protocol independent multicast (PIM-SM) projects at USC/ISI. In 2002, he received the NSF CAREER Award. His interests include network protocol design and analysis for mobile ad hoc and sensor networks, and mobility modeling.

\end{IEEEbiography}

\end{document}